\documentclass[twocolumn]{aastex701}

\usepackage{amsmath}
\usepackage[version=4]{mhchem}
\usepackage[T1]{fontenc}
\usepackage{multirow}
\usepackage{xcolor}

\definecolor{correction}{RGB}{0, 28, 254}

\begin{document}

\title{Theoretical determination of the binding energies of methanol and related species onto amorphous solid water ice}

\author[0009-0005-6138-492X]{Aneesa Ahmad}
\affiliation{School of Physics and Astronomy,University of Leeds, Leeds, LS2 9JT, UK}
\email{a.ahmad@leeds.ac.uk}

\correspondingauthor{C.~Walsh; \url{c.walsh1@leeds.ac.uk}}
\author[0000-0001-6078-786X]{Catherine Walsh}
\affiliation{School of Physics and Astronomy,University of Leeds, Leeds, LS2 9JT, UK}
\email{c.walsh1@leeds.ac.uk}

\correspondingauthor{S.~Vogt-Geisse; \url{stvogtgeisse@qcmmlab.com}}
\author[0000-0002-3102-1774]{Stefan Vogt-Geisse}
\affiliation{School of Physics and Astronomy,University of Leeds, Leeds, LS2 9JT, UK}
\affiliation{Departamento de Fisico-Quımica, Facultad de Ciencias Quımicas, Universidad de Concepción, Concepción, Chile}
\email{stvogtgeisse@qcmmlab.com}

\author[0009-0007-2699-6458]{Gabriela Silva-Vera}
\affiliation{Departamento de Fisico-Quımica, Facultad de Ciencias Quımicas, Universidad de Concepción, Concepción, Chile}
\email{stvogtgeisse@qcmmlab.com}

\author[0000-0003-0304-7931]{Felix Sainsbury-Martinez}
\affiliation{School of Physics and Astronomy,University of Leeds, Leeds, LS2 9JT, UK}
\email{f.sainsbury-martinez@leeds.ac.uk}


\begin{abstract}
The formation and survival of complex organic molecules (COMs) in cold interstellar environments depends on their interactions with icy dust grain surfaces. 
Methanol, a key COM detected in cold cores and protoplanetary disks, is believed to form on amorphous solid water (ASW) through surface reactions and reside 
there until it is desorbed into the gas phase. We present a theoretical study of the binding energies (BEs) of methanol and its photolysis-derived
species on ASW clusters by means of dispersion-corrected density functional theory (DFT) using  a refined protocol implemented in the
Binding Energy Evaluation Platform (BEEP).  Molecules capable of hydrogen bonding, such as \ce{H2O}, \ce{CH3OH}, HCOOH, and OH, exhibit high BEs and 
broad BE distributions that reflect the structural heterogeneity of the ASW surface. 
In contrast, weakly interacting volatiles including CO, \ce{CO2}, \ce{CH4}, and \ce{CH3} display narrower distributions dominated by dispersion interactions. Open-shell radicals such as \ce{CH2OH} and OH bind more strongly than HCO and \ce{CH3} due to their ability to form directional hydrogen bonds. 
Incorporation of our BEs into an astrochemical model, in conjunction with a recalculation of the pre-exponential factor using transition state theory, demonstrates the sensitivity of model results to the method of calculation of the grain-surface reaction rates. 
The new approach generally predicts a higher abundance of radicals on the ice that are key reactants for the formation of COMs when surface diffusion is assumed to be efficient.
These findings emphasize the importance of incorporating BEs that have been determined in a self-consistent manner into astrochemical models, and provide reliable theoretical benchmarks for species with limited experimental data.
\end{abstract}


\keywords{Astrochemistry (75) --- Astrophysical Dust Processes (99) --- Interstellar Dust (836) --- Protoplanetary Disks (1300) --- Surface Ices (2117)} 


\section{Introduction}
\label{sec:intro}

Stars and their surrounding planetary systems form within dark, dense molecular clouds in the interstellar medium. 
During the gravitational collapse of these clouds, dust and gas feed a protoplanetary disk, providing the material required to form a nascent planetary system \citep[e.g.,][]{VanDishoeck2020,Oberg2021}. 
The chemical evolution during star and planet formation is significantly influenced by icy mantles that form on dust grain surfaces \citep[e.g.,][]{Boogert2015}. 
These mantles consist of volatile species that form or condense onto dust grains at low temperatures ($\sim 10$~K) and densities ($\sim 10^3 - 10^4~\mathrm{cm}^{-3}$), creating an icy reservoir that can drive complex chemistry when exposed to the extreme conditions encountered as stars and their surrounding protoplanetary disks form \citep[e.g.,][]{Herbst2009,Jorgensen2020}.

Ice growth in the interstellar medium occurs predominantly through surface deposition and reactions of molecular and atomic species, with initial chemistry dominated by hydrogen atom addition reactions \citep[e.g.,][]{Linnartz2015}. 
For example, formation pathways that have been verified in the laboratory include the hydrogenation of oxygen atoms to form water \citep[\ce{H2O};][]{Hiraoka1998,Dulieu2010}, and hydrogenation of CO ice which yields species such as formaldehyde (\ce{H2CO}) and methanol \citep[\ce{CH3OH};][]{Hiraoka1994,Fuchs2009}. 
Larger, so-called complex organic molecules (COMs) can then form through recombination of reactive atoms and radicals produced in-situ in the ice via hydrogen abstraction reactions, or dissociation by UV or X-ray photons, or energetic electrons produced via the interaction of high energy photons and particles \citep[e.g.,][]{Bennett2007,Oberg2009,Chen2013,Chuang_2015}.

Methanol has long been recognised as a key feedstock for organic chemistry in star- and planet-forming environments \citep[e.g.,][]{Boogert2015,Oberg2016}, with its presence in the gas phase of protoplanetary disks recently confirmed through sensitive observations with the Atacama Large Millimeter/submillimeter Array \citep[ALMA; e.g.,][]{Walsh_2016,Booth2021,vanderMarel2021}. 
Methanol is also an important component of cometary ices  \citep[see, e.g.,][]{Lippi2024}, and so was likely abundant in the disk around the young Sun within which our Solar System formed. 
Observational studies summarised in \citet{Jorgensen2020} have further demonstrated the ubiquitous nature of methanol in star-forming environments, where it serves as a primary carrier of organic material. 
These findings highlight the importance of understanding the physical and chemical processes governing the formation and chemistry of methanol within icy mantles, as its partitioning between solid and gas phases depends on local physical conditions (temperature, density and strength of radiation) as well as the strength of the interactions of methanol and chemically-related species with various ice surfaces \citep[e.g.,][]{Das2008}.

Interstellar ices are thought to form layered structures where \ce{H2O} ice dominates the bulk, followed by layers rich in volatile organics such as CO, \ce{CO2}, and \ce{CH3OH} \citep[e.g.,][]{Boogert2015}. 
Recent JWST observations of interstellar ice composition along multiple sight lines have revealed a strong association between CO and \ce{CO2} ice at high densities where CO freezeout is advanced \citep{Smith2025}.
The composition and morphology of these layers reflect the evolving conditions in the parent molecular cloud, envelope, or disk, including the density, temperature, and strength and shape of any radiation fields that are present \citep{Herbst2009,Jorgensen2020,Oberg2021}. 
For instance, JWST spectra of ice absorption within protostellar envelopes show evidence for ice segregation upon heating to temperatures $\gtrsim 50$~K, with \ce{CO2} moving from a polar water-rich layer into one more mixed with \ce{CH3OH} \citep{Brunken2025}.
On the other hand, JWST observations of edge-on disks show that CO ice is present mainly in an apolar component mixed with \ce{CO2} \citep{Bergner2024,Bergner2026}.

Because methanol can act as both a hydrogen-bond (H-bond) donor and acceptor, it is particularly sensitive to the local ice environment. 
Experimental \citep[e.g.,][]{Collings2004} and theoretical \citep[e.g.,][]{Molpeceres2024}, studies have shown that the adsorption or binding energy (BE) of a species can vary significantly depending on whether it binds to \ce{H2O}-rich (i.e., polar) or CO/\ce{CO2}-rich (i.e., apolar) surfaces. 
Accurate determination of these BEs is crucial for astrophysical modelling, as it affects the predicted volatility, residence time, and potential reactivity of methanol and its related species in different regions of the disk or cloud \citep[e.g.,][]{Cuppen2017}.

Specific to the planet-forming environment, the BEs of key volatiles help define the radial extent and shape of snow lines in protoplanetary disks, marking where a given species transitions from the gas to the solid phase \citep[e.g.,][]{Oberg_2011}. 
The precise location of these snow lines influences other parameters of the disk important for its structure and evolution including the gas composition, dust grain growth, and ultimately planetary formation pathways, making accurate BE determinations essential for realistic models of planet-forming environments \citep[e.g.,][]{Kennedy2008,Drazkowska2017,Eistrup2018}. 
In particular, it is now proposed that the thermal desorption behaviour controlled by the BE affects the distribution of volatiles throughout the midplane of protoplanetary disks through the sublimation of ices from pebbles as they drift towards the star \citep{Piso_2015,Booth2019,Mah2023}.
Indeed, comprehensive chemical models rely on robust BE values to track the freeze-out and sublimation of molecules in protoplanetary disks, to set the efficiency of grain-surface chemistry, and ulitmately to constrain the chemical properties of emerging planetary systems \citep[e.g.,][]{Semenov2011,Furuya2014,Walsh2014}.

Previous research on BEs of key organic species to amorphous solid water (ASW) has involved both experimental and theoretical efforts. 
Temperature programmed desorption (TPD) experiments summarised in \citet[][]{Minissale2022} have provided constraints on the desorption kinetics using deposition techniques for a variety of molecules on ASW, revealing the significant influence of ice morphology and surface 
preparation conditions. 
From a theoretical perspective, force-field (MM) and quantum chemical (QM) simulations have systematically investigated the effects of cluster size, dispersion interactions, and binding site diversity. 
Such studies have employed a range of approaches to model water ice, including a large single cluster of 200 molecules \citep[e.g.,][]{Bariosco2025}, sets of 22-molecule amorphous clusters \citep[e.g.,][]{Bovolenta_2022}, small \ce{H2O} cluster (1-4) models \citep[e.g.,][]{Das_2018, Hendrix2024}, and force-field-based cluster representations \citep[e.g.,][]{Garrod2013,Karssemeijer2014}. 
Accurate BE predictions are shown to depend sensitively on both the underlying structural model and the theoretical method applied, with this dependence becoming particularly pronounced for strongly H-bonding species such as methanol. 
Despite these advances, BEs for radicals present a notable challenge. 
Experimental measurements are scarce due to the transient nature (and thus short lifetime) of radicals, and reliable theoretical predictions remain limited, highlighting the need for dedicated quantum chemical studies.

In this work, we present a comprehensive theoretical investigation of the BEs of methanol and its related species on amorphous water clusters. 
We focus on ASW as this is the dominant ice component in interstellar ices, and on methanol and its derived species given its importance as a feedstock for building chemical complexity in space. 
Whilst it is expected that \ce{CH3OH} and its derivatives will initially form in a CO-ice dominated layer, at elevated temperatures $\gtrsim 20$~K, such as internal to the CO snowline in protoplanetary disks, the binding environment will be dominated by water ice.
Most preceding works also assume water ice as the dominant ice component and this allows us to compare our results with the existing literature \cite[e.g.,][]{Wakelam_2017,Das_2018,Ferrero2020,Enrique-Romero2022,Hendrix2024,Sil_2024}.
Using high-level quantum chemical calculations, we determine precise BEs for a range of methanol-derived molecules and radicals. 
Our results aim to provide the astrochemical community with quantitatively robust parameters, with calculated BEs at high and consistent levels of theory, and the associated data to determine pre-exponential factors for computation of the desorption rates in astrochemical models, ultimately supporting efforts to model chemical evolution in star- and planet-forming regions and to interpret current and future observations of methanol and its complex organic derivatives. 
In Section~\ref{sec:comp} we describe our computational methodology, in Section~\ref{sec:res} we present our results, in Sections~\ref{sec:dis} and \ref{sec:chem-mod} we compare and contrast our results with those present in the literature and outlay the astrochemical consequences, and finally, in Section~\ref{sec:conc} we present our conclusions.


\section{Computational methodology}
\label{sec:comp}

\begin{figure}
    \includegraphics[width=0.45\textwidth]{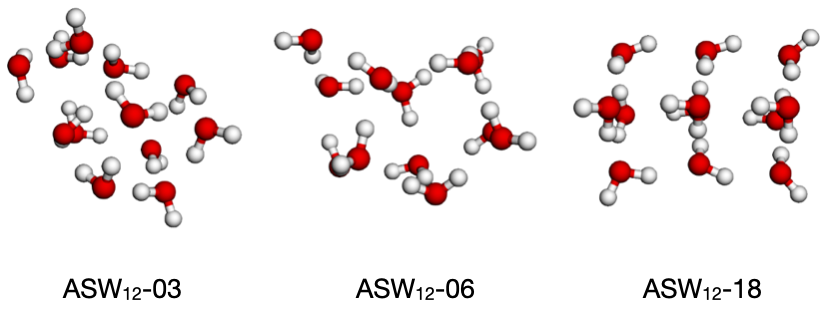}
    \caption{Three clusters which are representative of the sample of clusters used in this work. \label{fig:clusters}}
\end{figure}

The ASW surface ice model was adapted from the ``set-of-cluster" approach to obtain BE distributions using a similar methodology first introduced in \citet{Shimonishi2018} and \citet{bovolenta_2020}. 
We chose a cluster size of 12 water molecules 
to be able to use an accurate model chemistry for energies, gradients and Hessians, while at the same time having enough 
binding site diversity to obtain different binding modes and 
build a distribution of binding energies.
To obtain the ASW surface model, a 12-molecule 
cluster was thermalised at 300 K using \textit{ab-initio} molecular dynamics (MD) simulations at the BLYP/def2-SVP level of theory  \citep{PhysRevA.38.3098,PhysRevB.37.785,1989CPL...157..200M, 2005PCCP....7.3297W} to amorphise the system. 
From the high temperature trajectory, we selected 15 equally spaced frames and annealed the systems to 10~K. 
These 15 
clusters of 12 water molecules each span the set-of-cluster model
on which we sampled from binding sites. 
Figure~\ref{fig:clusters} shows three representative water-ice clusters for illustration.

\subsection{Computational framework}
\label{sec:comp1}

\begin{figure*}
    \centering
    \includegraphics[width=\textwidth]{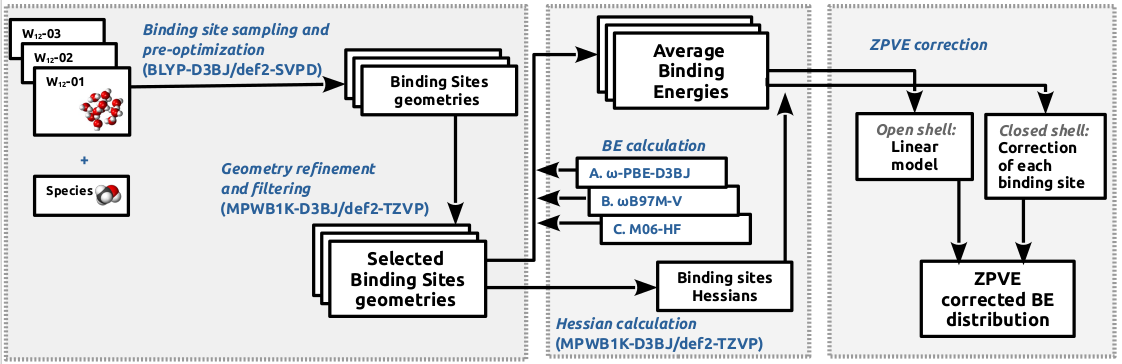}
    \caption{Three-workflow computational procedure used in this work for building 
    ZPVE-corrected BE distributions. \label{fig:flowchart}}    
\end{figure*}

The BEEP (Binding Energy Evaluation Platform) framework was utilised in this study, 
as first described in \cite{Bovolenta_2022}. The construction of the BE distributions
on a set-of-clusters model follows a structured three-step procedure:
i) sampling and geometry refinement,
ii) binding energy and Hessian computation, and iii) ZPVE (zero-point vibrational energy) corrected BE distribution assembly. Figure \ref{fig:flowchart}  shows a schematic diagram of the three workflows.  We used BEEP v0.6.0 for this study. The BEEP source code is publicly available\footnote{\url{www.github.com/svogt/beep}}.

\subsubsection{Levels of theory}
\label{sec:comp2}

Initial geometry optimisations for methanol and other closed-shell or stable species were performed at the BLYP-D3BJ/def2-SVPD \citep{1993JChPh..98.5648B, Rappoport2010, Hellweg2015} level of theory with Grimme's D3-BJ dispersion correction \citep{Grimme2010}. 
For radical species, we employed unrestricted BLYP calculations \citep[UBLYP/def2-SVPD;][]{1993JChPh..98.5648B, Rappoport2010, Grimme2010} using the same basis set and dispersion correction. 
We chose this level as it provides good initial geometries from the sampling round for further refinement.

After sampling, we refined these structures using MPWB1K-D3BJ/def2-TZVP
\citep{Zhao2004, Weigend2005, Grimme2010} for closed-shell species and
UMPWB1K/def2-TZVP-D3BJ \citep{Zhao2004, Weigend2005, Grimme2010} for open-shell 
(i.e., radical) systems. All DFT computations where done with Psi4 quantum 
chemistry software\footnote{\url{https://psicode.org/}} and the empirical dispersion correction was computed with dftd3 package.

\subsubsection{Binding energy formalism}
\label{sec:comp3}

The BE ($\Delta E_\mathrm{b}$) was obtained using a  basis set superposition error \citep[BSSE;][]{1970MolPh..19..553B} corrected 
electronic energy while also including the zero-point vibrational energy (ZPVE) contribution, 
\begin{equation}
\Delta E_\mathrm{b} = -(E_\mathrm{X + ice} - (E_\mathrm{X} + E_\mathrm{ice})    + \Delta_\mathrm{ZPVE}), 
\label{Equ:BEcomp}
\end{equation}
where X is the adsorbate species of interest. The negative 
sign in Eq.~\ref{Equ:BEcomp} is due to the convention of a positive 
definition of BE. 
$\Delta_{ZPVE}$ is defined as
\begin{equation}
 \Delta_{ZPVE} = E_{ZPVE,X+\text{ice}} - (E_{ZPVE,X} + E_{ZPVE,\text{ice}}).
\end{equation}
In the case of open-shell molecules, the ZPVE corrected BEs where obtained 
by scaling the non-ZPVE corrected molecules by the linear models obtained by 
relating ZPVE-corrected BEs with BEs on a single water cluster (see Appendix~\ref{sec:ZPVE}).

\subsubsection{Updated BEEP protocol}
\label{sec:comp4}

The BEEP protocol has recently been updated to incorporate additional refinements in the final BE calculation step. 
The final BE is  calculated as an average across three distinct levels of theory. These levels of theory were validated using benchmarks on small water clusters with one to three water molecules. The detailed results of the benchmark are given in Appendix~\ref{sec:bench}. 
Based on the benchmark results, the following levels of theory were chosen in this study for BE computation: $\omega$-PBE-D3BJ \citep{Perdew1996,Grimme_D3BJ}, $\omega$B97X-V \citep{10.1063/1.4952647} and M06-HF \citep{Zhao2008} with the def2-TZVP \citep{Weigend2005} basis set. 
For the more weakly bound species, \ce{CH4} and \ce{CH3}, we employed XC (exchange-correlation) functionals that performed best in the \ce{CH4} benchmark. 
These were B97-2-D3BJ \citep{B97-2, Grimme_D3BJ}, $\omega$B97M-V \citep{wb97mv} and TPSSh-D3BJ \citep{TPSSh_Staroverov_2003, Grimme_D3BJ}, with the same basis set. 
Overall, the standard deviation in the BEs computed using the three approaches was less than 200~K, which is comparable to the mean absolute error (MAE) of the different methods in the benchmark.

For all the binding sites of the closed-shell or stable molecules, we computed full Hessian matrices to determine the ZPVE corrections and filter structures with negative Hessian matrix eigenvalues.
Due to the high computational cost of calculating all Hessians for open-shell species, we adopted a linear model approach for the ZPVE correction based on a limited set of Hessian calculations  
(see Appendix~\ref{sec:ZPVE}). 

These ZPVE-corrected and BSSE-corrected BEs were then used to generate the final BE distribution for each ASW$_{12}$–X system, where X represents the adsorbate species under investigation.

\section{Results} 
\label{sec:res}
\begin{figure}
    \centering
    \includegraphics[width=\columnwidth]{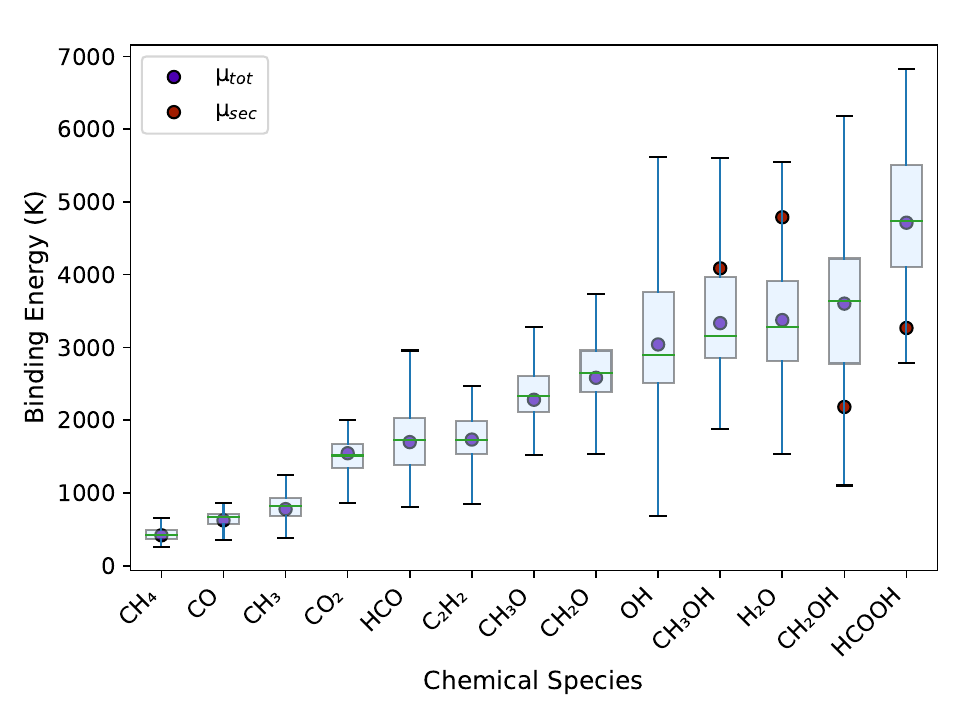}
    \caption{\centering Box plot representation of BE distributions, including $\mu_\mathrm{tot}$ (blue circle) and $\mu_\mathrm{sec}$ (red circle) values reported in this work.}
    \label{fig:all_data}
\end{figure}

We calculated BE distributions for thirteen species including methanol and its related species on ASW$_{12}$ clusters (see Fig.~\ref{fig:all_data}). 
Our analysis encompassed two molecular subsets; i) closed-shell or stable species (CO, \ce{CO2}, \ce{CH3OH}, \ce{H2O}, \ce{H2CO}, HCOOH, \ce{CH4}, and \ce{C2H2}),  and ii) open-shell radicals known to form through methanol photolysis (OH, \ce{CH3}, HCO, \ce{CH2OH}, and \ce{CH3O}).
For each species, we characterised the average primary ($\mu_\mathrm{tot}$) and secondary ($\mu_\mathrm{sec}$) binding modes (if present) as illustrated in Figs.~\ref{fig:coms-hbonds1}, \ref{fig:coms-hbonds2}, and \ref{fig:rads-BEs}. 
We compute the average binding energy ($\mu_\mathrm{tot}$ or $\mu_\mathrm{sec}$) and its associated error ($\sigma_\mathrm{tot}$ or $\sigma_\mathrm{sec}$) using Gaussian fits to the BE distributions using the same method as described in \citet{Bovolenta_2022}. 
We provide the full list of computed BEs for each system (\texttt{BE\_lists.tgz}), and all optimised structures (\texttt{all\_xyz.tgz}), in tar archives attached to the paper. 
All datasets are also available on Zenodo \footnote{\url{10.5281/zenodo.20764134}}.

Figures \ref{fig:all_data} and \ref{fig:coms-hbonds2} reveal that volatile species such as CO and \ce{CH4} exhibit narrower BE distributions. 
This suggests that these species interact relatively weakly and uniformly with the surface, resulting in less variability across different adsorption sites. 
Less volatile species such as \ce{H2O}, \ce{CH3OH} and HCOOH show broader distributions, reflecting the presence of multiple binding environments, including sites capable of forming strong H-bonds (see Figs.~\ref{fig:all_data} and \ref{fig:coms-hbonds1}). 
These trends highlight the importance of considering a range of BEs when modelling strongly interacting molecules on ASW ices. 
Building upon these general observations, we next examine the BEs of the closed-shell and open-shell species, highlighting site-specific interactions.

\subsection{Closed-shell species} 
\label{sec:res1}

\begin{figure*}
    \centering
    \includegraphics[width=0.7\textwidth]{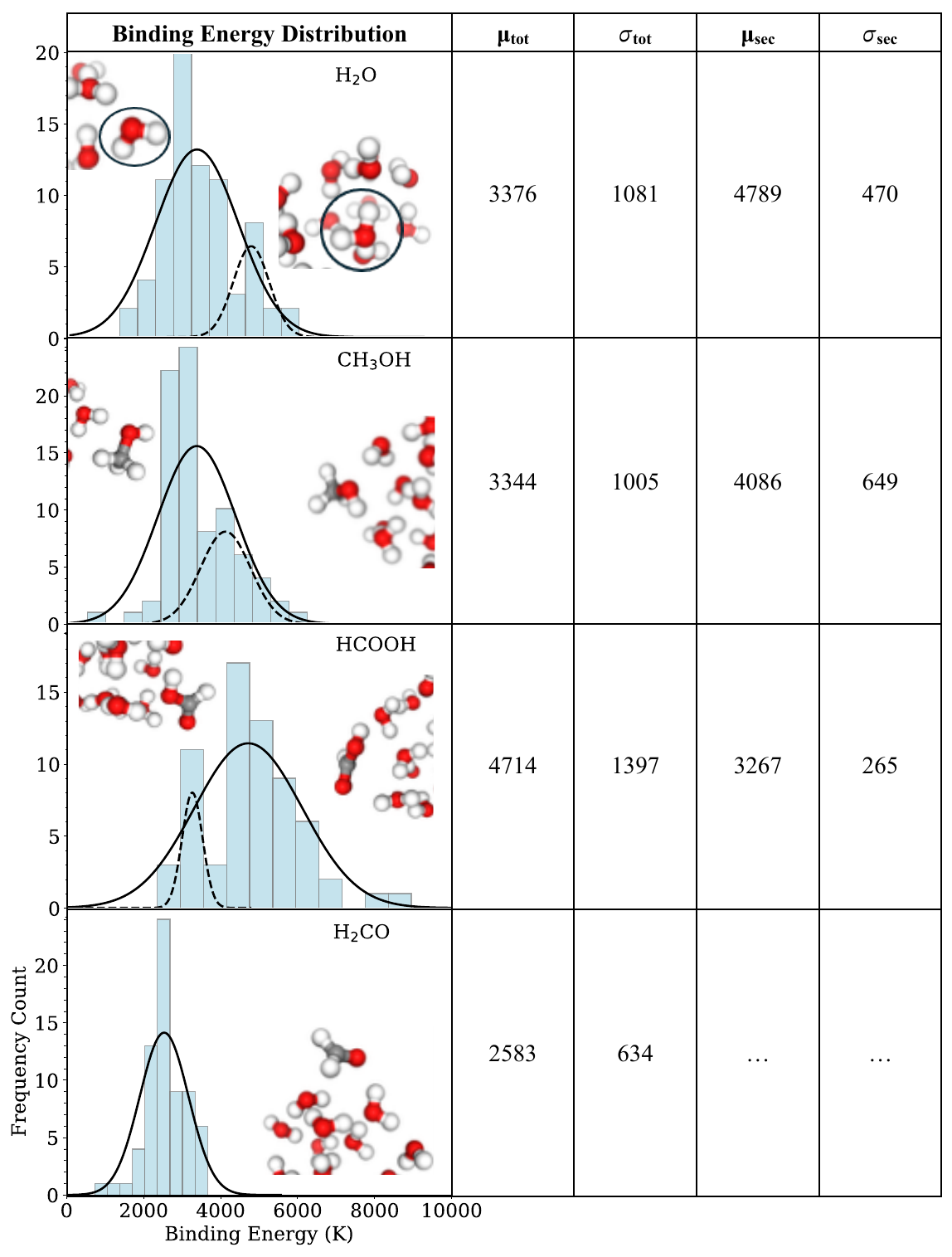}
    \caption{\centering BE distributions and binding modes for hydrogen-bonded molecules (\ce{H2O}, \ce{CH3OH}, HCOOH, and \ce{H2CO}).Two fits are shown in the figure: the fit to the full BE data ($\mu_{tot}$, $\sigma_{tot}$) and a fit to the secondary binding mode ($\mu_{sec}$, $\sigma_{sec}$), if present. 
    Units are in K.
    An example of stationary point geometries of the different binding modes is shown in the inset images. The circled water molecules in the top panel identifies those considered as bound or adsorbed to the ASW$_{12}$ cluster.}
    \label{fig:coms-hbonds1}
\end{figure*}

\begin{figure*}
    \centering
    \includegraphics[width=0.525\textwidth]{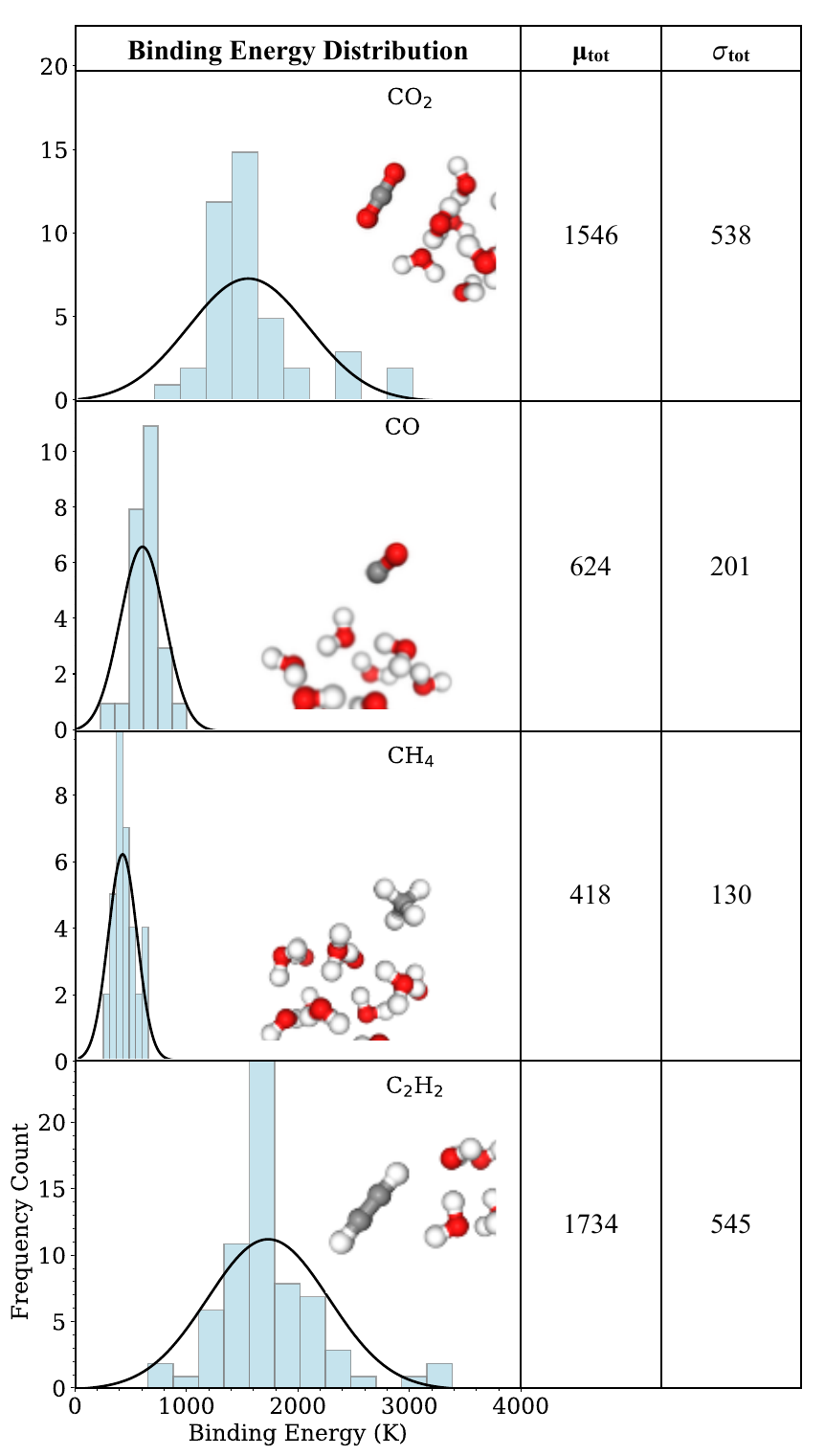}
    \caption{\centering BE distributions and binding modes for molecules bound by dispersion (\ce{CO2}, CO, \ce{CH4}, and \ce{C2H2}). Units are in K.
    An example of stationary point geometries of the  different binding modes is shown in the inset images.}
    \label{fig:coms-hbonds2}
\end{figure*} 

We calculated the BEs for eight closed-shell species which we divided into two groups.  
One group which have broad BE distributions sometimes with multiple binding modes (\ce{H2O}, \ce{CH3OH}, HCOOH, and \ce{H2CO}; see Fig.~\ref{fig:coms-hbonds1}), and for which the primary mode of binding is through H-bonds. 
In the second group, dispersion interactions 
are significant for binding (\ce{CO2}, CO, \ce{CH4}, and \ce{C2H2}; see Fig.~\ref{fig:coms-hbonds2}).  
For the species that have more than one binding mode, we report two values of the BE: $\mu_\mathrm{tot}$ is the average binding energy computed from a Gaussian fit to the entire distribution which is dominated by the primary binding mode, and $\mu_\mathrm{sec}$ is the value from a restricted fit over the secondary binding mode only.
For species with a single binding mode, we report $\mu_\mathrm{tot}$ only.

\ce{H2O} exhibits two distinct binding modes (see Fig.~\ref{fig:coms-hbonds1}). 
The stronger secondary mode, $\mu_\mathrm{sec}$ ($4789\pm470$~K), corresponds to configurations where \ce{H2O} forms two H-bonds with the surface of the ASW$_{12}$ clusters. 
The weaker primary mode, $\mu_\mathrm{tot}$ ($3376\pm1081$~K), reflects binding sites with dangling-OH (d-OH) configurations with reduced intermolecular contact with the surface of the cluster. 
This is in agreement with previous research that has also shown that the presence of d-OH groups can lead to weaker interactions with adsorbed species, resulting in lower BEs \citep{ Perera2009, Hsieh2011, Henkelman2016,Michoulier_2019}.

Similarly, \ce{CH3OH} also demonstrates two characteristic binding modes. 
The primary mode, $\mu_\mathrm{tot}$ ($3344\pm1005$~K), emerges when the \ce{-OH} moiety forms a single H-bond, and the \ce{-CH3} moiety weakly interacts with the surface of the cluster. 
The secondary more stable bidentate binding mode, $\mu_\mathrm{sec}$ ($4086\pm649$~K), is a result of two strong H-bonds forming with the surface. 
Similar to \ce{H2O}, the weaker of the two binding modes for \ce{CH3OH} also involves d-OH groups, as shown in Fig.~\ref{fig:coms-hbonds1}.

HCOOH was the final closed-shell or stable species that we determined to have two binding regimes. 
The stronger binding mode, $\mu_\mathrm{tot}$ ($4714\pm1397$~K), is a result of the formation of two H-bonds forming with the surface of the ASW$_{12}$ clusters. 
HCOOH exhibits high BEs because it can both donate and accept hydrogen bonds via its OH and \ce{C=O} groups, thus resulting in a strong interaction of HCOOH with the clusters. 
The secondary mode, $\mu_\mathrm{sec}$ ($3267\pm265$~K), also forms two H-bonds with the surface of the clusters. 
However, the configuration is not parallel to the surface, resulting in a weaker surface interaction with HCOOH. 
The donor and acceptor behaviour of HCOOH is reflected in the broad distribution of BEs, highlighting the extended range of BEs for this adsorbate molecule.

Finally, \ce{H2CO} exhibits only one peak in its BE distribution indicating
the presence of a dominating binding mode in which the carbonyl oxygen 
receives a H-bond while the \ce{-CH2} establishes an H-bond with a water oxygen 
acceptor. However, as has been shown in our previous work, 
the carbonyl oxygen can accept more than one H-bond thus increasing the 
binding energy with the surface \citep{Bovolenta2024}. 

The BE of the molecules in the second group is notably lower because they lack strong hydrogen bonding interactions with the water molecules 
on the cluster surface and are not considerably polarizable (see Fig.~\ref{fig:coms-hbonds2}). 
\ce{CH4} with zero dipole moment and \ce{CO} with a very small dipole moment (0.122~D)
stand out as the molecules with the lowest BE due to the absence of polarization and therefore a significant portion of the 
BE is due to weak dispersion interactions.  On the other hand \ce{CO2} which has a polarized electron density towards the oxygen atoms, and 
\ce{C2H2} with a polarizable $\pi$-bond, display stronger binding to the water clusters. Due to the lack of H-bonds and a single binding motif, 
the molecules in this group display a single BE distribution. 

\subsection{Radicals} 
\label{sec:res2}

\begin{figure*}
    \centering
    \includegraphics[width=0.7\textwidth]{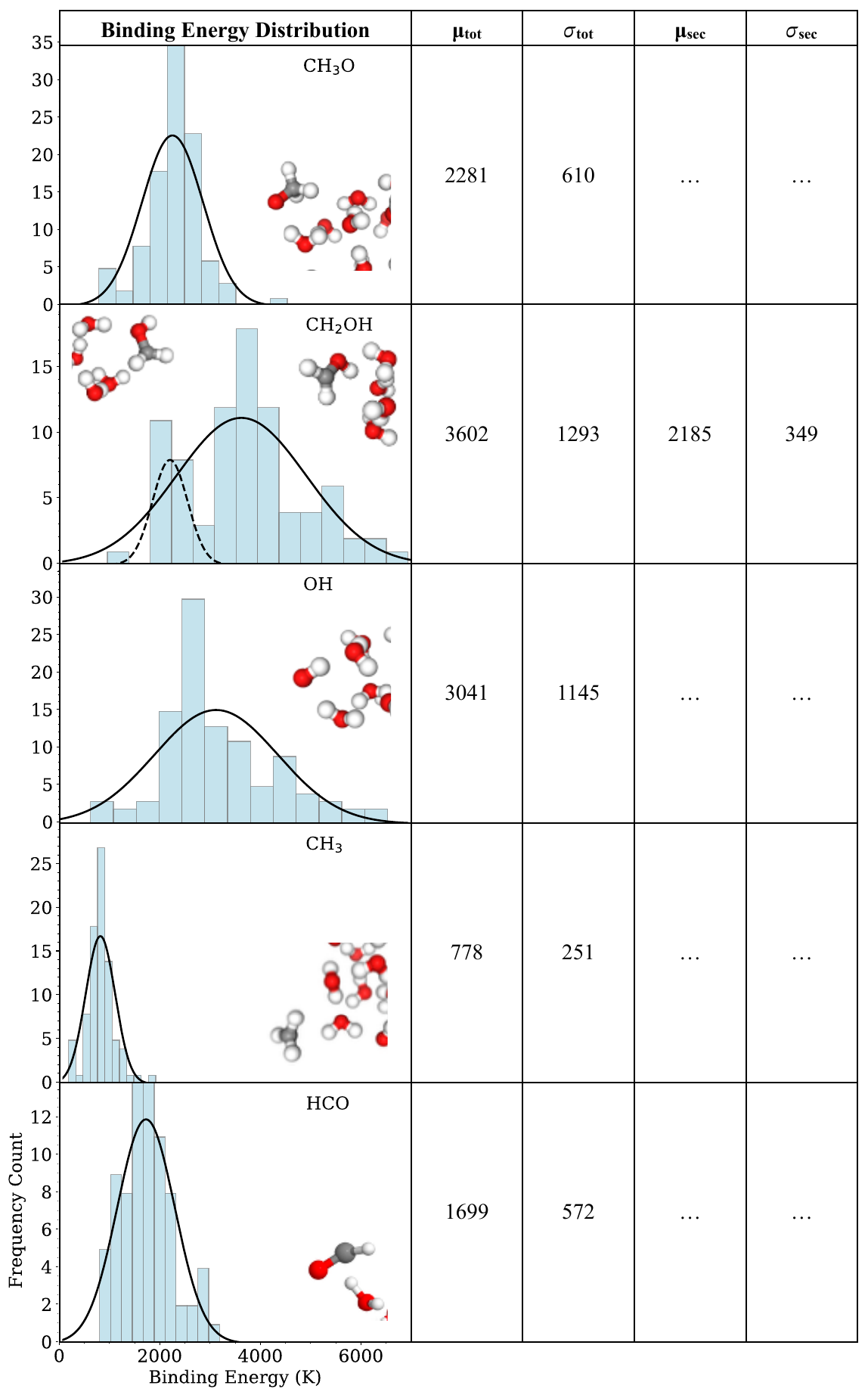}
    \caption{BE distributions and binding modes for radical species (\ce{CH3O}, \ce{CH2OH}, OH, \ce{CH3}, and HCO).  Two fits are shown in the figure: the fit to the full BE data ($\mu_{tot}$, $\sigma_{tot}$) 
    and a fit to the secondary binding mode ($\mu_{sec}$, $\sigma_{sec}$), if present. Units are in K.
    An example of stationary point geometries of the 
    different binding modes is displayed in the inset images.}
    \label{fig:rads-BEs}
\end{figure*}

The BE values and distributions of open-shell radical species, namely \ce{CH3O}, \ce{CH2OH}, OH, \ce{CH3}, and HCO are shown in Fig.~\ref{fig:rads-BEs}. 
Open-shell species demonstrate binding characteristics distinct from those of their closed-shell counterparts 
as a result of the unpaired electron that alters their hydrogen-bonding patterns and preferred adsorption orientations.

The \ce{CH3O} radical exhibits its primary interactions through its oxygen moiety, similar to \ce{CH3OH}'s \ce{-OH} group, but with altered surface orientation. 
\ce{CH3O} only accepts an H-bond due to the lack of an \ce{-OH} group, yielding an average BE of $2281\pm610$ K, lower than the average BE for methanol ($\mu_\mathrm{tot} = 3419\pm139$~K). 

Amongst the radicals, \ce{CH2OH} is the only one that displays dual-binding behaviour, mediated by its OH moiety and radical carbon centre. 
This results in distinct binding modes with varying degrees of H-bond participation. 
The stronger binding mode, $\mu_\mathrm{tot}$ ($3602\pm1293$~K), has a strong interaction of its OH moiety with the ASW$_{12}$ cluster surfaces. 
The secondary mode, $\mu_\mathrm{sec}$ ($2185\pm349$~K), has a d-OH bond resulting in a weaker overall interaction of the \ce{CH2OH} radical, similar to our findings for \ce{CH3OH} and \ce{H2O}. The slightly overall higher BE of  \ce{CH2OH} than that of \ce{CH3OH} is due to the less bulky
\ce{-CH2} moiety compared to the methyl group in \ce{CH3OH}, that allows the radical species to interact more efficiently with the water 
molecules on the surface.

The OH radical demonstrates strong binding
as it can versatilely insert itself into the water H-bond network, resulting in a high BE of $3041\pm1145$~K. 
On the other hand, \ce{CH3} ($778\pm251$~K) lacks strong H-bonding capabilities, interacting primarily through dispersion forces, producing BE 
behaviour reminiscent of its closed-shell analogue \ce{CH4}, but modified by the radical center that enables
a slightly stronger interaction. 
The binding energy for the HCO radical is $1699\pm572$~K, with binding predominantly occurring through an H-bond between the oxygen and the ASW surface. HCO binds noticeably weaker to the ASW surface compared
to other O-bearing radicals, and forms a weaker H-bond to the surface. This might
be due to the more linear geometry that hinders the establishment 
of an efficient donor-acceptor H-bond interaction with the surface.

These findings demonstrate that electronic structure modifications resulting from radical formation can significantly influence adsorption geometries and energetics.

\section{Discussion}
\label{sec:dis}

\subsection{Comparison with reported experimental binding energies}
\label{sec:dis1}

\begin{figure}[h]
    \centering
    \includegraphics[width=\columnwidth]{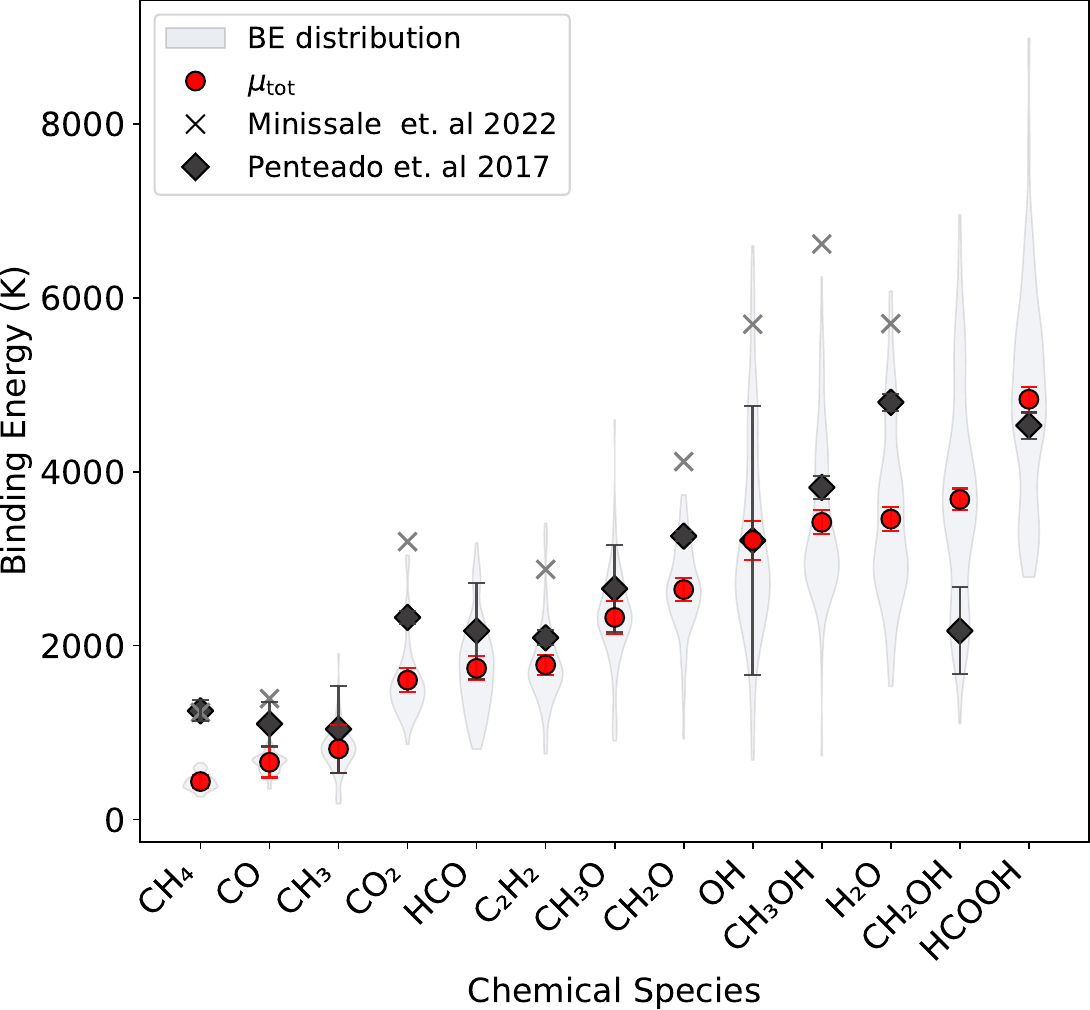}
    \caption{Comparison of the average BEs for this work (red circles) compared to experimental values compiled in the literature (grey crosses for \citealt{Minissale2022} and black diamonds for \citealt{Penteado2017}). The full BE distribution from this work is shown as violin plot for each species.
    See Table~\ref{tab:experimental_BEs} for further details and references to original data. 
    }
    \label{fig:exptl-comp}
\end{figure}

\begin{deluxetable*}{lccccccccc}
\tablecaption{BEs (K) compiled by \citet{Penteado2017} and \citet{Minissale2022} for species bound to ASW compared to those computed in this work from Gaussian fits to the distributions ($\mu_\mathrm{tot}$). 
Also shown are the maximum BEs (K) from our distributions ($BE_\mathrm{max}$) and the uncertainty which arises from the chosen DFT functionals ($\sigma_\mathrm{comp}$).}
\label{tab:experimental_BEs}
\tabletypesize{\scriptsize}
\tablewidth{0pt}
\tablehead{
     & \multicolumn{3}{c}{This work} & & \multicolumn{2}{c}{\citet{Minissale2022}} & & \multicolumn{2}{c}{\citet{Penteado2017}} \\
    \cline{2-4} \cline{6-7} \cline{9-10} 
    & $\mu_\mathrm{tot}$ & $BE_\mathrm{max}$ & $\sigma_\mathrm{comp}$ & & Recommended$^a$ & \colhead{Average$^b$} & & \colhead{Recommended$^c$} & 
    \colhead{Experimental$^d$}
}
\startdata
\multirow{2}{*}{\ce{H2O}} & \multirow{2}{*}{$3376\pm1081$} & \multirow{2}{*}{6040} & \multirow{2}{*}{138} & & \multirow{2}{*}{5705} & \multirow{2}{*}{5640} && \multirow{2}{*}{$4800\pm100$} & $4815\pm15$, {$5070\pm50$, 5600,} \\
 &  &  &  &&     &          &&   & {$4799\pm96$, $5930\pm240$, 4800} \\
\ce{CH3OH} & $3344\pm1005$ & 6204 & 139 && 6621    & 5000         && $3820\pm135$  & \nodata \\
\ce{HCOOH} & $4714\pm1397$ & 8930 & 144 && \nodata & \nodata      && $4532\pm150$  & \nodata \\
\ce{H2CO}  & $2583\pm634$  & 3711 & 130 && 4117    & 3260         && $3260\pm60$   & $3260\pm60$\\
\ce{CO2}   & $1546\pm538$  & 3021 & 141 && 3196    & $2105\pm902$ && $2267\pm70$   & 2236 -- 2346\\
\ce{CO}    & $624\pm201$   & 1019 & 175 && 1390    & \nodata      && $1100\pm250$  & 863 -- 1307, 1420\\
\ce{CH4}   & $418\pm130$   & 650  & 72 && 1232    & $1368\pm68$  && $1250\pm120$  & 1370 \\
\ce{C2H2}  & $1743\pm545$  & 3386 & 113 && 2877    & $3000\pm220$ && $2090\pm85$   & \nodata \\
\ce{CH3O}  & $2281\pm610$  & 4567 & 192 && \nodata & \nodata      && $2655\pm500$  & \nodata \\
\ce{CH2OH} & $3602\pm1293$ & 6914 & 126 && \nodata & \nodata      && $2170\pm500$  & \nodata \\
\ce{OH}    & $3041\pm1145$ & 6559 & 226 && 5698    & 4600         && $3210\pm1550$ & 1656 -- 4760\\
\ce{CH3}   & $791\pm251$   & 1896 & 180 && \nodata & \nodata      && $1040\pm500$  & \nodata \\
\ce{HCO}   & $1699\pm572$  & 3162 & 137 && \nodata & \nodata      && $1355\pm500$  & \nodata \\
\enddata
\tablecomments{The listed uncertainties, $\sigma_\mathrm{comp}$, are the standard deviation of the different DFT functionals used to compute the BE. 
$^a$Recommended BE values from \citet{Minissale2022}. 
$^b$Average value from compilation of experimental data by \citet{Minissale2022} with error bar stated where provided \citep[data from][]{Sandford1990,Speedy1996,Fraser2001,Wakelam_2017,noble2012,Noble_2012,Edridge2013,Smith2016,Behmard2019,Dulieu2013,Miyazaki2020}. 
$^c$Recommended BE values from \citet{Penteado2017} with recommended error bar also stated.
$^d$Experimental values compiled and used by \citet{Penteado2017} with errors bars where provided \citep[data from][]{Acharyya2014,Fraser2001,Brown2007,Collings2015,Dulieu2013,noble2012,Noble_2012,He2014,HeVidali2014}.}
\end{deluxetable*}

To assess the consistency of our theoretically-calculated BEs, we contrast them with the datasets compiled by \citet{Minissale2022} and \citet{Penteado2017}, as summarized in Table~\ref{tab:experimental_BEs}, and illustrated in Figure~\ref{fig:exptl-comp}.  
In Table~\ref{tab:experimental_BEs} we also show the uncertainty which arises from the choice of DFT functional ($\sigma_\mathrm{comp}$), and the BE for the strongest binding site identified for each species ($BE_\mathrm{max}$). 
In Fig.~\ref{fig:exptl-comp} we only plot data for which there are available experimental values. 

\citet{Minissale2022} compiled experimentally determined BEs obtained primarily via TPD experiments, where BEs are derived from the data using the Polanyi-Wigner equation. 
These experimental values inherently reflect surface dependent effects such as sub-monolayer coverage, surface morphology, and heterogeneity in binding sites. 
In contrast, \citet{Penteado2017} compiled BEs from a mixture of experimental reports and theoretical estimates to be used in 
astrochemical gas-grain modelling to explore the sensitivity of model results on the assumed suite of BEs. 
Their work derives the BEs from TPD experiments where data are available, by scaling the binding energy relative to that for \ce{H2O} using 
the ratio of the desorption temperatures as measured from TPD curves. 

The experimental BEs reported in Table~\ref{tab:experimental_BEs} and illustrated in Fig.~\ref{fig:exptl-comp} are, in general, larger than those derived from our theoretical calculations. 
For example, the average BE for \ce{H2O} in our simulations is $3376$ K, whereas the experimentally reported values range between 4800--5930 K (see references listed in Table~\ref{tab:experimental_BEs}).  

Surface morphology and coverage  contribute significantly to the observed differences. 
As an example, experiments by \citet{Noble_2015} measuring the thermal desorption of \ce{O2} and CO from a sub-monolayer mixture reveal that, under sub-monolayer conditions, molecules preferentially occupy the strongest available binding sites, which in turn leads to a TPD peak that is shifted to higher temperatures and consequently a higher BE \citep[see also, e.g.,][]{Fraser2001,Smith2016}. 
We also list the binding energies for the strongest binding sites, $BE_\mathrm{max}$, identified in our calculated distributions. 
In some cases the strongest identified binding sites lies closer to those derived from TPD experiments (e.g., CO and \ce{H2CO}) than the peak value we derived from a Gaussian fit to the data ($\mu_\mathrm{tot}$).  
For most species, however, the strongest binding site is significantly larger than the values derived from experiment.
This is most notable for the species for which H-bonds play an important role in the binding (e.g., \ce{H2O}, \ce{CH3OH}, HCOOH, \ce{CH2OH}, and OH).

The discrepancies between our results and experimental data are particularly pronounced for volatile species like CO and \ce{CH4}, consistent with the difficulty of accurately probing weak physisorption interactions in experiments. However, these weak interactions could play an important role in the  desorption and diffusion patterns of volatiles species in cold interstellar environments \citep[see also, e.g.,][]{Bariosco2025}.
Conversely, for strongly-bound species such as the OH radical and HCOOH, our calculated BEs show better agreement with the experimental determinations. 
These findings emphasise the importance of accounting for surface heterogeneity when interpreting BEs derived from laboratory measurements, and they highlight the complementary role of theoretical modelling in advancing our understanding of grain-surface interactions and chemistry.

\subsection{Comparison with reported theoretical binding energies}
\label{sec:dis2}

\begin{figure*}
    \centering
    \includegraphics[width=0.9\textwidth]{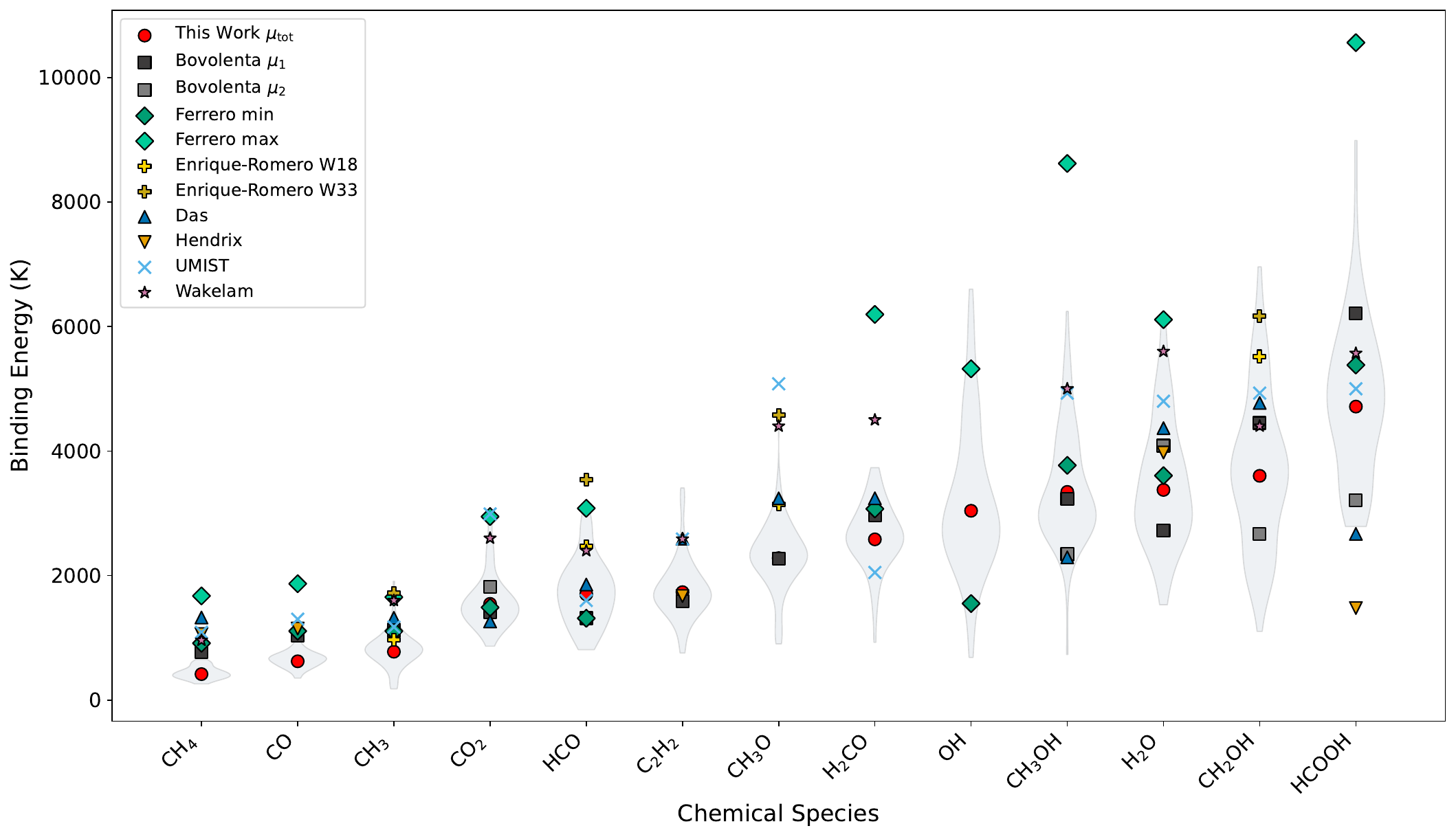}
    \caption{\centering Comparison of theoretically-determined BEs from this work compared to those from the literature. The full BE distribution from this work is shown as violin plot for each species. See Table~\ref{tab:theoretical_BEs} for further details and references to original data.}
    \label{fig:theory-comp}
\end{figure*}

\begin{deluxetable*}{lccccccccccccccc}
\caption{\centering{Theoretically-determined and database BEs from the literature compared to those computed in this work.}\label{tab:theoretical_BEs}}
\tablewidth{0pt}
\tabletypesize{\scriptsize}
\tablehead{
\colhead{Species} & 
\multicolumn{2}{c}{This Work} & 
\multicolumn{2}{c}{Bovolenta$^a$} & 
\multicolumn{2}{c}{Ferrero$^b$} & 
\colhead{Sil$^c$} & 
\colhead{Das$^d$} & 
\colhead{Hendrix$^e$} & 
\multicolumn{2}{c}{Enrique-Romero$^f$} &
\multicolumn{2}{c}{Wakelam$^g$} & 
\colhead{UMIST$^h$} \\
\colhead{} & 
\colhead{$\mu_\mathrm{tot}$} & 
\colhead{$\mu_\mathrm{sec}$} & 
\colhead{$\mu_1$} & 
\colhead{$\mu_2$} & 
\colhead{min} & 
\colhead{max} & 
\colhead{} & 
\colhead{} & 
\colhead{} & 
\colhead{W18} & 
\colhead{W33} & 
\colhead{$\mu_1$} & 
\colhead{$\mu_2$} & 
\colhead{}
}
\startdata
\ce{H2O}   & $3376\pm1081$ & $4789\pm470$ & 2725    & 4087    & 3605    & 6111    & \nodata & 4368    & 3976    & \nodata & \nodata & 4600    & \nodata & 4800 \\
\ce{CH3OH} & $3344\pm1005$ & $4086\pm649$ & 3235    & 2344    & 3770    & 8618    & \nodata & 2293    & \nodata & \nodata & \nodata & 4500    & 5100    & 4930 \\
HCOOH      & $4714\pm1397$ & $3267\pm265$ & 6207    & 3206    & 5382    & 10,559  & \nodata & 2670    & 1479    & \nodata & \nodata & \nodata & \nodata & 5000 \\
\ce{H2CO}  & $2583\pm634$  & \nodata      & 2970    & \nodata & 3071    & 6194    & \nodata & 3242    & \nodata & \nodata & \nodata & 5100    & \nodata & 2050 \\
\ce{CO2}   & $1546\pm538$  & $2615\pm303$ & 1408    & 1819    & 1489    & 2948    & \nodata & 1263    & \nodata & \nodata & \nodata & 3100    & \nodata & 2990 \\
CO         & $624\pm201$   &  \nodata     & 1035    & \nodata & 1109    & 1869    & \nodata & \nodata & 1150    & \nodata & \nodata & 1300    & \nodata & 1150 \\
\ce{CH4}   & $418\pm130$   &  \nodata     & 773     & \nodata & 914     & 1674    & \nodata & 1327    & 1061    & \nodata & \nodata & 800     & \nodata & 1090 \\
\ce{C2H2}  & $1734\pm545$  &  \nodata     & 1590    & \nodata & \nodata & \nodata & \nodata & 2593    & 1671    & \nodata & \nodata & 2600    & \nodata & 2587 \\
\ce{CH3O}  & $2281\pm610$  &  \nodata     & 2274    & \nodata & \nodata & \nodata & \nodata & 3240    & \nodata & 3139    & 4582    & 4700    & \nodata & 5080 \\
\ce{CH2OH} & $3602\pm1293$ & $2185\pm349$ & 4451    & 2670    & \nodata & \nodata & \nodata & 4772    & \nodata & 5521    & 6170    & 3900    & 5800    & 5084 \\
OH         & $3041\pm1145$ & \nodata      & \nodata & \nodata & \nodata & \nodata & 4287    & \nodata & \nodata & \nodata & \nodata & 3300    & 5300    & 2850 \\
\ce{CH3}   & $778\pm251$   & \nodata      & 1109    & \nodata & 1654    & \nodata & \nodata & 1322    & \nodata & 974     & 1720    & 2500    & \nodata & 1175 \\
HCO        & $1699\pm572$  & \nodata      & 1317    & \nodata & 1315    & 3081    & \nodata & 1857    & \nodata & 2466    & 3536    & 2300    & 2700    & 1600 \\
\enddata
\tablecomments{
$^a$\citet{Bovolenta_2022}, 
$^b$\citet{Ferrero2020},
$^c$\citet{Sil_2024},
$^d$\citet{Das_2018}, 
$^e$\citet{Hendrix2024}, 
$^f$\citet{Enrique-Romero2022},
$^g$\citet{Wakelam_2017},
$^h$\citet{McElroy2013}. \\
The mean and secondary BEs, $\mu_\mathrm{tot}$ and $\mu_\mathrm{sec}$ (or $\mu_\mathrm{tot}$ if a single binding mode), are reported for this work. 
BEs in Columns 2–12 were theoretically determined, while those in Column 13 are sourced from the compilation provided by the UMIST Database for Astrochemistry \citep[{\sc Rate}12 version;][]{McElroy2013}. 
We report the values from \citet{Enrique-Romero2022} for both 18-water molecule (W18) and 33-water molecule (W33) clusters.}
\end{deluxetable*}

Our study adopts the BEEP protocol, following the approach outlined by \citet{Bovolenta_2022}. 
Similar to \citet{Bovolenta_2022}, we explore the distribution of BEs by examining all unique adsorption sites on a set of water clusters; however, our use of a set of ASW$_{12}$ clusters (as opposed to ASW$_{22}$) and our multi-level DFT approach (e.g., WPBE-D3BJ, $\omega$B97X-V, M06-HF) introduce some differences in the final calculated BEs. 
Figure \ref{fig:theory-comp} visualises the impact of the methodological differences of our theoretically-calculated BEs to the literature. 
Since the set-of-clusters model ASW explicitly accounts for 
the diversity of binding sites accessible on the surface, it is not surprising that our distributions encompass the range of 
theoretical values reported in previous studies.

Table~\ref{tab:theoretical_BEs} shows how our computed values compare to \cite{Bovolenta_2022}. In general the values fall within the range of our
BE distributions; however the mean values are marginally different due to both the different cluster size, and the more refined 
level of theory employed in this study. 
For the volatiles, the discrepancies are larger due to the lack of application 
of the ZPVE correction in data  presented in \cite{Bovolenta_2022}. 
Comparison 
of the non-ZPVE-corrected binding energies reveals the effect of cluster size,
particularly for CO, for which binding energy distribution data are available 
at comparable DFT levels of theory. Here we report an average
BE of 930 K. When using a 22-water molecule cluster as in \cite{Bovolenta_2022} the average value increases to 1030 K using the M05-2x/def2-tzvp level of theory for the binding site optimization and $\omega$-PBE-D3BJ/def2-tzvp for the BSSE corrected BEs, reflecting a 10\% increase in the average BE value. 
In \citet{Bovolenta2025} the same CO BE distribution was computed 
using a machine learning potential (trained on MPwB1K-D3BJ/def-tzvp level of theory) on several large 500 water molecule surfaces. 
This very realistic surface provided an average BE of 1350 K again increasing 
the BE relative to the smaller cluster systems. We expect that a similar model size effect will be also observed for other molecules like \ce{CH3OH} and \ce{HCO}. Even though 
the average BE values stemming from the distribution on ASW$_{12}$ clusters using the set-of-clusters approach provides a systematically lower BE values than larger 
systems, it still samples the same binding modes that one can find on the 
more realistic surface providing accurate relative BEs among 
the species in this study and for the employed surface model.

Other variations stemming from cluster shape and sampling strategies become evident when comparing our results with those of \citet{Ferrero2020} and \citet{Enrique-Romero2022} who used a similar ASW model. 
In particular, \cite{Ferrero2020} examined up to eight binding sites per species on a periodic amorphous solid water 
(ASW) slab of 60 water molecules.  Their larger and continuous ice representation yielded in average higher BEs, 
likely due several binding sites being in a cavity resulting in the 
interaction with a greater number of water molecules, and a better 
representation of H-bond cooperativity effects in the larger system. 
In contrast, our sampling procedure explores all unique binding sites,
thus capturing the heterogeneity of binding configurations leading to our 
reported $\mu_\mathrm{tot}$ and $\mu_\mathrm{sec}$ 
values that effectively encompass both weaker and stronger interactions.

Methodological choices also play a key role in reported discrepancies. 
High-level quantum chemical techniques, such as the CCSD(T) \citep{Cizek1966OnTC} approach employed by \citet{Hendrix2024} for small water clusters (N= 1-4), often 
yield benchmark-level accuracy for specific configurations. 
However, their high computational cost limits the number of adsorbates and binding sites that can be treated, possibly overlooking broader site 
variability effects. Our dispersion-corrected DFT approach mitigates this limitation by enabling the treatment of larger ASW$_{12}$ clusters 
while still accounting for dispersion effects. 
As a result, our BEs for larger or strongly H-bound species, like HCOOH, depart substantially from smaller-cluster CCSD(T) values (e.g., 4714~K and 3267~K versus 
1479~K reported by \citealt{Hendrix2024}), but capture bidentate H-bonding motifs more comprehensively.

Studies by \citet{Sil_2024} and \citet{Das_2018} further highlight how differences in cluster size, ice morphology, 
and computational methods influence 
the calculated BEs of interstellar species on water ices. 
\citet{Sil_2024} report significantly higher BEs for OH (4287~K) compared to our mean value of 3041~K, partly due to their use of a larger 
ASW$_{20}$ cluster. However, their BE distribution appears to be derived from a targeted selection of high-binding-energy sites, potentially introducing a sampling bias that over-represents strong adsorption environments. 
In contrast, our BEEP approach samples the surface more uniformly, yielding a broader and more physically representative distribution of BEs. 
Meanwhile, \citet{Das_2018} report over- or under-estimations relative to ours, which can be attributed to the consideration of a single binding 
site for their study. 
On the other hand, the values reported by \cite{Wakelam_2017} are systematically higher than our results, as they are derived from a linear model that related BEs computed on water monomer to experimental TPD BE values. 

Recently \cite{Bariosco2025} obtained a full BE distribution for \ce{CH3OH} on a large 200 water molecule 
cluster that includes several concave regions using ONIOM\footnote{Our own N-layered Integrated molecular Orbital and molecular Mechanics, see e.g., \citet[][]{Chung2015} for a review.} embedding methods. 
They obtained a mean value of 4255 K for the BE of \ce{CH3OH} which is 800 K higher than 
the value reported here. 
The spread in the BE values is also 500 K wider than the one obtained in this study. 
This is expected for molecules that can form multiple hydrogen bonds to the ASW surface 
since the presence of nano-pores in larger clusters can enhance the strength of the hydrogen bonding and hence the BE.

\subsection{Comparison with data collated by the UMIST database}
\label{sec:dis3}

Our results can also be placed in the context of widely used astrochemical databases such as the UMIST Database for Astrochemistry \citep{McElroy2013} which compiled BEs from a mix of experimental data, older theoretical calculations, and empirical estimates. 
While these databases provide essential inputs for astrochemical modelling, their values often lack the uniform level of refinement that comes from a single, high-level protocol such as the approach highlighted in BEEP.

For instance, our two binding regimes for \ce{H2O} ($\mu_\mathrm{tot} = 3376$ K and $\mu_\mathrm{sec} = 4789$ K) lie below the UMIST (4800~K) value. 
Similar patterns emerge for \ce{CH3OH}, where our bidentate H-bonding results (3344~K and 4086~K) are lower than the 4930~K value listed in UMIST. 
The differences are particularly pronounced for weaker adsorbates like CO, where our low BE (624~K) contrasts with that compiled by UMIST (1150 K).
Taking a set of binding energies from disparate sources (i.e., a mixture of theoretical and experimental data) that follow different methodologies and experimental design will generate a set of binding energies not fully consistent with each other.  
What likely matters most for grain-surface chemistry are the {\em relative} binding energies of species which are competitors in grain-surface reactions, and those which are important for setting the locations of snowlines. 
The approach adopted here allows such a set of binding energies to be determined.

\section{Astrochemical modelling} 
\label{sec:chem-mod}

To test the impact of using a consistently-derived suite of binding energies in an astrochemical model, we simulate the chemistry of a dark cloud with a temperature of 10~K (for both gas and dust), an \ce{H2} number density of $10^{4}$~cm$^{-3}$, a visual extinction of 10~magnitudes, and a cosmic-ray ionisation rate of $1.3\times 10^{-17}$~s$^{-1}$.  
We use the chemical network and model from \citet{Walsh2015} and adopt initial elemental abundances from \citet{McElroy2013}. 

The BE is used in the model in two ways.  
Firstly, for the thermal desorption rate, $k_\mathrm{des}$, 
\begin{equation}
    k_\mathrm{des} = \nu \exp\left(\frac{-\Delta E_b}{T}\right),
\end{equation}
where the BE ($\Delta E_b$) is in units of K and $\nu$ is the characteristic frequency of the bound species, often also referred to as the ``pre-exponential factor". 
The BE is also used to estimate the size of the barrier to hopping, and hence controls the rate of diffusion of ice species across the surface, and by extension, the grain-surface reaction rates.  
The rate of thermal hopping, $k_\mathrm{hop}$, is similarly
\begin{equation}
    k_\mathrm{hop} = \nu \exp\left(\frac{-\chi \Delta E_b}{T}\right),
\end{equation}
where $\chi$ is a scaling factor that sets the ratio of the barrier to diffusion to that for desorption. 
Lower values of $\chi$ are typically assumed for diffusion across the ice surface ($0.3 - 0.5$), whereas higher values ($\approx 0.7$) are assumed for diffusion of molecules within the bulk ice mantle in so-called three-phase models, in which the chemistry in the mantle is computed separately from that on the surface \citep[see, e.g.,][]{Garrod2013,Chen2024}.
Here, we run a two-phase gas-grain chemical model (i.e., assuming that chemistry occurs on the ice surface only) where we include a large gas-phase network (the base for which is {\sc Rate}12; \citealt{McElroy2013}), freeze-out, thermal desorption, photodesorption (driven by cosmic-ray-induced UV), reactive desorption (with an efficiency of 1\%), two-body grain-surface reactions (via the Langmuir-Hinshelwood mechanism), and ice photodissociation (driven by cosmic-ray-induced UV).  
The ratio of the barrier to diffusion to that for desorption is an uncertain parameter not yet well constrained by experiment nor theory, except in a handful of cases (see the discussions in \citealt{Chen2024} and \citealt{Ligterink2025}). 
To explore the effects on the results on the assumed ratio for the barrier to diffusion to that of desorption, $\chi$, we run models using a value of 0.3 (faster diffusion) and 0.5 (slower diffusion).  
We note here that the diffusion of H and \ce{H2} is controlled solely by quantum tunnelling at 10~K \citep[e.g.,][]{Kuwahata2015}; hence, the rates of H and \ce{H2} diffusion is the same across all models.

We run models adopting different sets of BEs. 
The first is one in which we use the BEs for the species of interest here as compiled by the {\sc Rate}12 version of the UMIST Database for Astrochemistry (see Table~\ref{tab:experimental_BEs}) and for which we adopt the harmonic oscillator approximation to compute the pre-exponential factor, $\nu$ \citep{1992ApJS...82..167H},
\begin{equation}
    \nu_\mathrm{HO} = \sqrt{\frac{2\,n_s k_{B}\Delta E_b}{\pi^2 m}},
\label{equ:HO}
\end{equation}
where $n_s$ is the number density of surface sites, $k_{B}$ is Boltzmann's constant, $\Delta E_b$ is the BE in units of K, and $m$ is the mass of the bound species. 
This equation typically gives values of the order of $10^{12}$~s$^{-1}$ for all species. 

In a second model, we replace the binding energies of the species of interest here with those derived from the fit to our BE distributions ($\mu_\mathrm{tot}$) and we recalculate the pre-exponential factor, $\nu_\mathrm{TST}$, using transition state theory (TST). 
As outlined in \citet{Minissale2022},
\begin{equation}
    \nu_\mathrm{TST} = \frac{k_{B}T}{h}q_\mathrm{tran,2D}q_\mathrm{rot,3D},
\label{equ:TST}
\end{equation}
where $h$ is Planck's constant and $q_\mathrm{tran,2D}$ and $q_\mathrm{rot,3D}$ are the 2D translational and 3D rotational partition functions for an adsorbed species, respectively \citep[see][for further details]{Minissale2022}. 

In the computation of $\nu$ using TST, the moments of inertia, $I_x$, of the adsorbed species are required and we compute them in this work for geometries optimized at the MPWB1K-D3BJ/def2-tzvp level of theory.
In Table~\ref{tab:Iandnu} we list the symmetry numbers and moments of inertia for each species, as well as a comparison of the characteristic frequencies computed using the harmonic oscillator approximation, $\nu_\mathrm{HO}$ (Eq.~\ref{equ:HO}), and those computed using transition state theory, $\nu_\mathrm{TST}$ (Eq.~\ref{equ:TST}), at a temperature of 10~K which is that assumed in our chemical model.
In these calculations, and in the chemical model, we assume a number density of surface sites of $6\times10^{14}$~cm$^{-2}$. 
For some species, $\nu_\mathrm{TST}$ is slightly lower than or similar to $\nu_\mathrm{HO}$ which is the case for \ce{H2O}, OH, \ce{CH4}, and \ce{CH3}; however, for most species, $\nu_\mathrm{TST}$ is around 1 to 2 orders of magnitude larger.
For these species, the thermal desorption rates and thermal hopping rates will be faster, compared with those computed using the harmonic oscillator approximation which is traditionally assumed in astrochemical models \citep[see the detailed discussion in][]{Minissale2022}.

\begin{deluxetable*}{lccccccc}
    \caption{\centering{Symmetry numbers, moments of inertia, $I_x$, and  pre-exponential factors, $\nu$, for the species considered in this work. \label{tab:Iandnu}}}
\tablewidth{0pt}
\tabletypesize{\small}
\tablehead{\colhead{Species} & \colhead{Symmetry number} & \colhead{$I_a$} & \colhead{$I_b$} & \colhead{$I_c$} && \colhead{$\nu_\mathrm{HO}$} & \colhead{$\nu_\mathrm{TST}$(10~K)} \\ 
\cline{3-5} \cline{7-8}
 & & \multicolumn{3}{c}{[amu~\AA$^2$]} & & \multicolumn{2}{c}{[s$^{-1}$]}}
\startdata
\ce{H2O}   & 2  & 0.60 & 1.16  & 1.75  && $1.4\times 10^{12}$ & $5.4\times 10^{11}$ \\
\ce{CH3OH} & 1  & 3.87 & 19.89 & 20.61 && $1.0\times 10^{12}$ & $6.9\times 10^{13}$ \\
\ce{HCOOH} & 1  & 6.41 & 40.83 & 47.24 && $1.0\times 10^{12}$ & $2.8\times 10^{14}$ \\
\ce{H2CO}  & 1  & 1.76 & 12.63 & 14.39 && $9.3\times 10^{11}$ & $2.9\times 10^{13}$ \\
\ce{CO2}   & 2  & 0.00 & 42.08 & 42.08 && $5.9\times 10^{11}$ & $7.8\times 10^{13}$ \\
\ce{CO}    & 1  & 0.00 & 8.51  & 8.51  && $4.7\times 10^{11}$ & $2.0\times 10^{13}$ \\
\ce{CH4}   & 12 & 3.15 & 3.15  & 3.15  && $5.1\times 10^{11}$ & $4.1\times 10^{11}$ \\
\ce{C2H2}  & 2  & 0.00 & 13.98 & 13.98 && $8.2\times 10^{11}$ & $1.5\times 10^{13}$ \\
\ce{CH3O}  & 3  & 3.16 & 17.67 & 17.79 && $8.6\times 10^{11}$ & $1.8\times 10^{13}$ \\
\ce{CH2OH} & 1  & 4.72 & 17.35 & 20.15 && $1.1\times 10^{12}$ & $6.9\times 10^{13}$ \\
\ce{OH}    & 1  & 0.00 & 0.88  & 0.88  && $1.3\times 10^{12}$ & $1.3\times 10^{12}$ \\
\ce{CH3}   & 3  & 1.74 & 1.74  & 3.48  && $7.3\times 10^{11}$ & $8.8\times 10^{11}$ \\
\ce{HCO}   & 1  & 0.68 & 10.99 & 11.68 && $7.7\times 10^{11}$ & $1.5\times 10^{13}$ \\
\enddata
\tablecomments{$\nu_\mathrm{HO}$ refers to the pre-exponential factor calculated using the harmonic oscillator approximation \citep[e.g.,][]{1992ApJS...82..167H}, whereas $\nu_\mathrm{TST}$ refers to that calculated using transition state theory \citep[e.g.,][]{Minissale2022}.}
\end{deluxetable*}

\begin{figure*}
\centering
    \includegraphics[width=\textwidth]{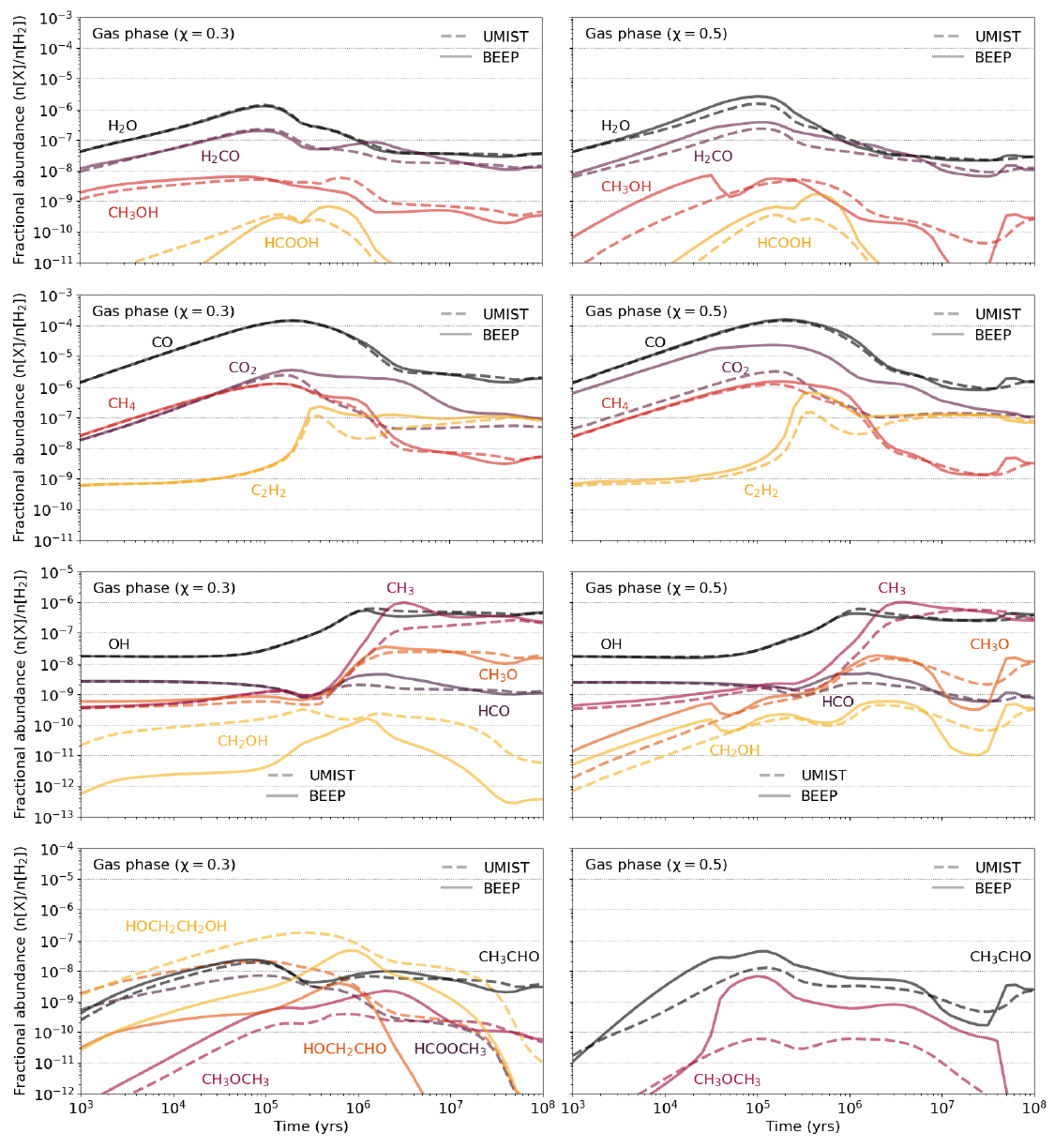}
    \caption{Fractional abundances of gas-phase species as a function of time in a single-point model of a dark cloud. Shown are results for a model adopting BEs from the UMIST Database for Astrochemistry and the harmonic oscillator approximation for the characteristic frequency ``UMIST''; dashed lines).  Also shown are results for a model adopting the BEs computed for the species considered in this work for which the characteristic frequencies were also computed using transition state theory (``BEEP''; solid lines). The left- and right-hand plots show results adopting a ratio of the diffusion barrier to desorption barrier, $\chi$, of 0.3 and 0.5, respectively. \label{fig:model-results-gas}}
\end{figure*}

\begin{figure*}
\centering
    \includegraphics[width=\textwidth]{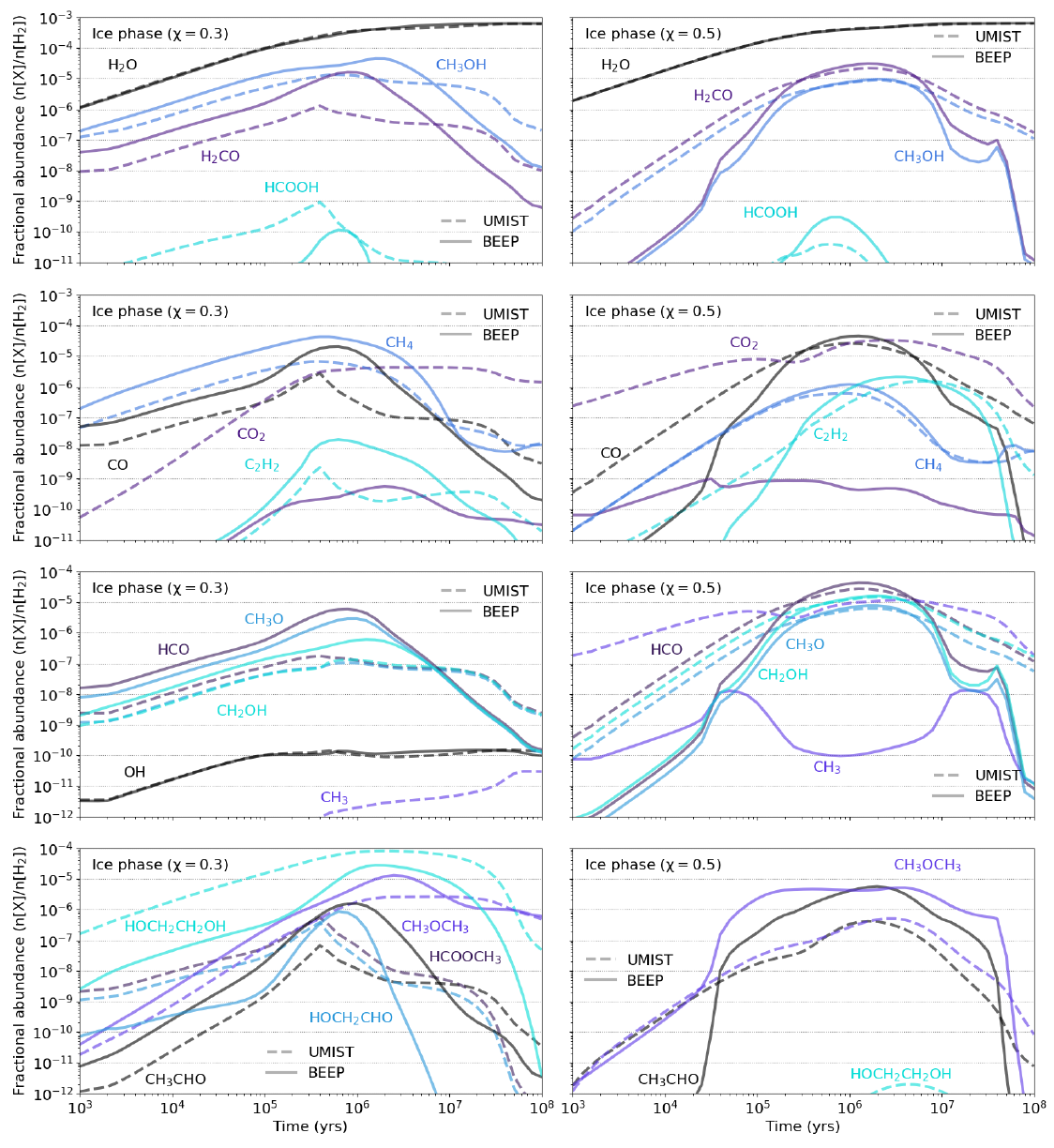}
    \caption{Same as Fig.~\ref{fig:model-results-gas} for ice-phase species. \label{fig:model-results-ice}}
\end{figure*}

Figures~\ref{fig:model-results-gas} and \ref{fig:model-results-ice} show the results from a single-point model of a dark cloud using the BEs for the species considered here from those compiled by the UMIST database in dashed lines (``UMIST''), and those from our calculations, including calculating the pre-exponential factor using transition state theory, in solid lines (``BEEP'').
We plot the fractional abundances for all species considered here in both the gas (Fig.~\ref{fig:model-results-gas}) and ice (Fig.~\ref{fig:model-results-ice}) phase, for two values of $\chi$, the ratio of the barrier to diffusion to that for desorption ($\chi=0.3$ and 0.5 in the left- and right-hand panels, respectively). 
We also plot the fractional abundances of larger complex organic species which have been postulated to be directly formed from the radicals we consider in this work, and thus are chemically related to methanol \citep[see, e.g.,][]{Oberg2009}.

For closed-shell stable species we see only small differences in their gas-phase abundances comparing the results from both approaches. 
For instance, the results for water gas and ice are almost identical regardless of the assumed BEs and the adopted value for $\chi$ \citep[see also][]{Chen2024}. 
This is because in our model, water ice formation is dependent on the rate of hydrogenation via quantum tunnelling. 
We see a slight decrease and increase by a factor of a few in gas-phase \ce{H2CO} and \ce{CH3OH} at late times ($\gtrsim 10^5$~yrs), respectively, when using the BEEP values for both values of $\chi$.
There is also an increase in the abundance of gas-phase \ce{CO2} by around one to two orders of magnitude (between $\sim 10^{5}$ and $\sim 10^{7}$~yrs when $\chi=0.3$), and an increase in the abundance of \ce{C2H2} by a factor of a few beyond a few times $10^{5}$~years. 
The BEEP model assuming slower diffusion ($\chi =0.5$) predicts a noticeably higher gas-phase \ce{CO2} abundance, reaching a peak of a few times $10^{-5}$ with respect to \ce{H2}.
Regarding the gas-phase radical species, the abundances of OH gas and ice are practically identical in all models and there are only small differences for the gas-phase abundances of \ce{CH3O} and HCO over the full model timescale when comparing the UMIST results with the BEEP results. 
For the slower diffusion model ($\chi=0.5$), the abundance of gas-phase \ce{CH3O} and \ce{CH2OH} decrease and increase respectively, compared with the faster diffusion model ($\chi=0.3$).
The abundance of gas-phase \ce{CH3} is higher by a factor of a few just beyond $10^{6}$~yrs when using the BEEP approach for both values of $\chi$.
The abundance of gas-phase \ce{CH2OH} is more sensitive to the assumed diffusion barrier. 
Its abundance is lower by up to two orders of magnitude in the BEEP model than those computed using the UMIST values when $\chi=0.3$; however, when diffusion is assumed to be slower, the differences between the BEEP and UMIST models are more minor. 

There is a lot more variation between the models for the large complex organic molecules (COMs).   
These species are generally {\em only} formed efficiently on the grain surfaces and are released into the gas phase via non-thermal desorption. 
A key difference between the models with fast ($\chi=0.3$) and slow ($\chi=0.5$) diffusion is the absence of methyl formate (\ce{HCOOCH3}),  glycol aldehyde (\ce{HOCH2CHO}) and ethylene glycol (\ce{HOCH2CH2OH}) in the gas phase when diffusion is slow, indicating the sensitivity of the formation rate of these species to the assumed diffusion rate. 
In the fast diffusion case, the BEEP model predicts a lower abundance of ethylene glycol (\ce{HOCH2CH2OH}), glycolaldehyde (\ce{HOCH2CHO}), and methyl formate (\ce{HCOOCH3}), and slightly more dimethyl ether (\ce{CH3OCH3}), in the gas phase. 
Both model approaches predict similar abundances of gas-phase acetaldehyde, \ce{CH3CHO}, suggesting the importance of gas-phase chemistry in setting its abundance. 
The decrease in \ce{CH2OH}-containing species is likely due to the higher binding energy adopted in the BEEP approach. 
For the \ce{CH3O}-containing species, this radical is less weakly bound, and thus more mobile, and there is a competition between radical-radical recombination with hydrogenation to reform \ce{CH3OH}. 
In the slow diffusion model, the BEEP approach generally predicts a higher abundance of \ce{CH3OCH3} and \ce{CH3CHO} than when using the UMIST approach, with the abundance of \ce{CH3OCH3} in the BEEP model higher by between one and two orders of magnitude.

For the ice species (Fig.~\ref{fig:model-results-ice}), in general, the BEEP approach predicts higher abundances of \ce{H2CO}, \ce{CH3OH}, \ce{CH4}, CO, and \ce{C2H2} when diffusion is fast. 
When diffusion is slow, the picture is more complex. 
The rates of formation of the ice species differ when using both approaches (indicated by the gradient); however, similar abundances of \ce{H2CO}, \ce{CH3OH}, \ce{CH4}, CO, and \ce{C2H2} are reached between $\sim 10^{5} - 10^{7}$~years. 
The BEEP approach predicts a significantly lower abundance of \ce{CO2} ice regardless of the assumed barrier to diffusion, indicating increased competition between those species formed via reactions with CO on the ice surface.

Similarly, for the fast diffusion case, the BEEP approach generally predicts higher abundances of radicals in the ice phase, with the exception of \ce{CH3} which has negligible abundances over the whole timescale ($\lesssim 10^{-13}$).  
This could be due to the lower binding energy for \ce{CH3} predicted in this work which will increase its hopping and desorption rates and thus reduce its residency time on the grain. 
When diffusion is slow, the rates of formation of the radicals on the ice is faster (i.e., traced by the steeper increase in the abundance with time), when using the BEEP approach; however, again, similar abundances of HCO, \ce{CH3O}, and \ce{CH2OH} are obtained between $\sim 10^{5} - 10^{7}$~years. 
\ce{CH3} is the exception here once again, with the UMIST model predicting a significantly higher abundance in the ice (reaching a peak of $\sim 10^{-5}$) versus that in the BEEP model ($\lesssim 10^{-8}$).

For the COMs, the ice abundances mimic those in the gas phase in the fast diffusion case, with \ce{HOCH2CH2OH}, \ce{HOCH2CHO}, and \ce{HCOOCH3} having less efficient formation, and \ce{CH3OCH3} and \ce{CH3CHO} having slightly more efficient formation.  
These latter two species can be formed from association reactions involving the \ce{CH3} radical, and their synthesis could be increased in efficiency by the increased mobility of this species on the ice surface. 
For the slow diffusion case, only \ce{CH3OCH3} and \ce{CH3CHO} are efficiently formed in the ice phase, with the BEEP model predicting higher abundances than the UMIST model. 
The assumption of slow diffusion suppresses the formation of \ce{HOCH2CHO} and \ce{HOCH2CH2OH} in both models.
The disappearance of methyl formate (\ce{HCOOCH3}), glycolaldehyde (\ce{HOCH2CHO}), and ethylene glycol (\ce{HOCH2CH2OH}) in the slow-diffusion model indicates that the predicted abundances of these species are highly sensitive to the adopted diffusion-to-binding energy ratio, $\chi$. 
This highlights the need, in future work, to compute or constrain species-specific diffusion barriers, or distributions of diffusion barriers on realistic ice surfaces, rather than relying only on a fixed fraction of the binding energy.

We discuss here some caveats to our modelling approach. 
Firstly, we use a two-phase model so that we do not distinguish fully the chemistry occurring in the bulk ice from that occurring in the surface.  
However, we do mediate the surface diffusion and reaction rates by allowing only two monolayers of ice to be chemically ``active". 
Nonetheless, use of a three-phase or multi-phase gas-grain chemical model may mute somewhat the differences seen here between the slow and fast diffusion models.  
Further, the models that we have run are not fully internally consistent in the approach to computing the thermal desorption and hopping rates. 
We recompute these {\em only} for the species considered here, and adopt the traditional approach (i.e., assuming harmonic oscillator) for all other species in the network.  
Hence, these models should be treated only as illustrative of the differences that can be expected when using a more appropriate approach for a sub-set of species in the network and we make no judgement on the accuracy of the model results when compared with, for example, observations of interstellar ices. 


\section{Conclusions}
\label{sec:conc}

In this work, we have provided an accurate and self-consistent \textit{ab initio} set of BE distributions for 13 methanol and methanol-derived
closed- and open-shell species adsorbed on ASW, to be used in astrochemical models. Using an improved  BEEP protocol applied to the 
set-of-cluster ASW$_{12}$ model, we used multiple curated  DFT functionals for the calculation of BEs which put together yielded 
a BE distribution for each species. All binding sites geometries were obtained at the DFT level of theory paired with large basis 
sets; ZPVE corrections were included for all BEs resulting in highly accurate values for the employed model surface. 

Species, such as \ce{H2O}, \ce{CH3OH}, HCOOH, and OH, exhibit 
broader BE distributions, attributed to their ability to form multiple H-bonds with the ice surface. 
In contrast, volatile and weakly interacting species such as CO, \ce{CO2}, \ce{CH4}, and \ce{CH3} 
display narrower distributions. H-bonding leads to higher BEs, particularly for molecules
possessing donor and acceptor functional groups (e.g., \ce{HCOOH} and \ce{CH3OH}). 
We also obtained DFT level BE distributions for key radicals for surface chemistry such as \ce{CH3O}, \ce{CH2OH}, 
HCO, and OH, with the \ce{CH2OH} and OH radicals exhibiting the highest BE amongst open-shell species due to 
their strong H-bonding capabilities and compact form that allows for a more efficient interaction with the surface 
water molecules. Furthermore, CH${_3}$O and CH${_2}$OH radicals showed binding characteristics reminiscent of their methanol
parent molecule, whereas HCO and CH${_3}$, with limited H-bonding ability, had weaker BEs. 

We used the BEs computed here in conjunction with the recalculation of the pre-exponential factors, $\nu$, using transition state theory, 
in a single-point chemical model of a dark cloud. 
We compared the results with a model in which we use the UMIST binding energies and calculate $\nu$ using the harmonic oscillator approximation and assumed two different values for the ratio of the barrier for diffusion to that for desorption.
The abundances of both \ce{H2O} and OH are very similar when using either approach, and there is a slightly increased efficiency in ice-phase \ce{H2CO} and \ce{CH3OH} formation when using the BEEP values for the case of fast diffusion. 
The BEEP approach with fast diffusion also showed less efficient synthesis of \ce{CH2OH}-containing ice species such as \ce{HOCH2CH2OH} and \ce{HOCH2CHO}, and more efficient synthesis of \ce{CH3O}-containing ice species such as \ce{CH3OCH3} and \ce{CH3CHO}.  
There were several extreme responses in the chemistry, including a drop in the abundance of both \ce{CH3} and \ce{CO2} in the ice in the BEEP models, as well as \ce{HCOOCH3} in both phases.  
Further, the formation of \ce{HOCH2CHO} and \ce{HOCH2CH2OH} ice and gas were suppressed when diffusion was assumed to be slow in both models.
This indicates that the chemistry, even at a low temperature of 10~K, is sensitive to the assumed BEs and to the method used to compute the characteristic frequencies. 
These results highlight the need for BE calculations for all ice species present in astrochemical networks to produce a self-consistent suite of data.

Our findings highlight the importance of incorporating realistic BEs into astrochemical models, particularly for species with strong surface interactions, and for radicals which are poorly constrained experimentally. 
Future work will extend these results to a larger range of species, and to the predictions of BEs for mixed ASW-methanol clusters and will introduce the binding behaviours of \ce{CO2} ice clusters.


\begin{acknowledgements}
\label{sec:ack}
A.~A. and F.~S.~M. acknowledge support from UK Research and Innovation (grant numbers MR/T040726/1 and MR/Z00029X/1). 
C.W.~acknowledges financial support from the Science and Technology Facilities Council and UK Research and Innovation (grant numbers ST/X001016/1, MR/T040726/1 and MR/Z00029X/1).  S.V-G acknowledges financial support from the VRID research grant 2022000507INV.
\end{acknowledgements}


\bibliography{papers.bib}
\bibliographystyle{aasjournalv7}


\appendix
\section{ZPVE Models} 
\label{sec:ZPVE}

\begin{figure*}
    \centering
    \includegraphics[width=0.75\textwidth]{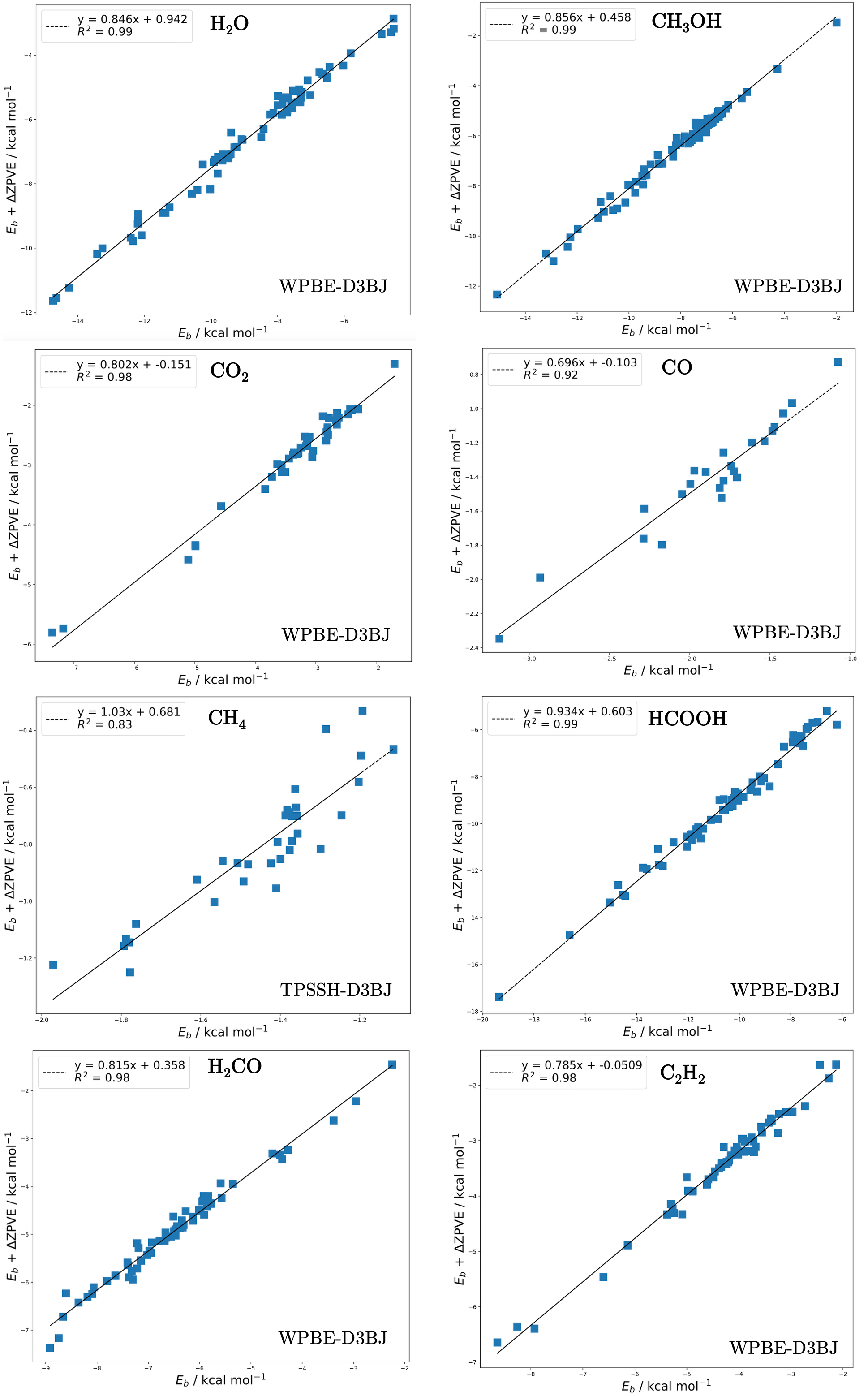}
    \caption{Complete ZPVE models for stable molecules at the WPBE-D3BJ level of theory. Note that \ce{CH4} illustrated a poor linear fit, hence this model was calculated at the TPSSH-D3BJ level of theory.}
    \label{fig:ZPVE_coms}
\end{figure*}

\begin{figure*}
    \centering
    \includegraphics[width=0.75\textwidth]{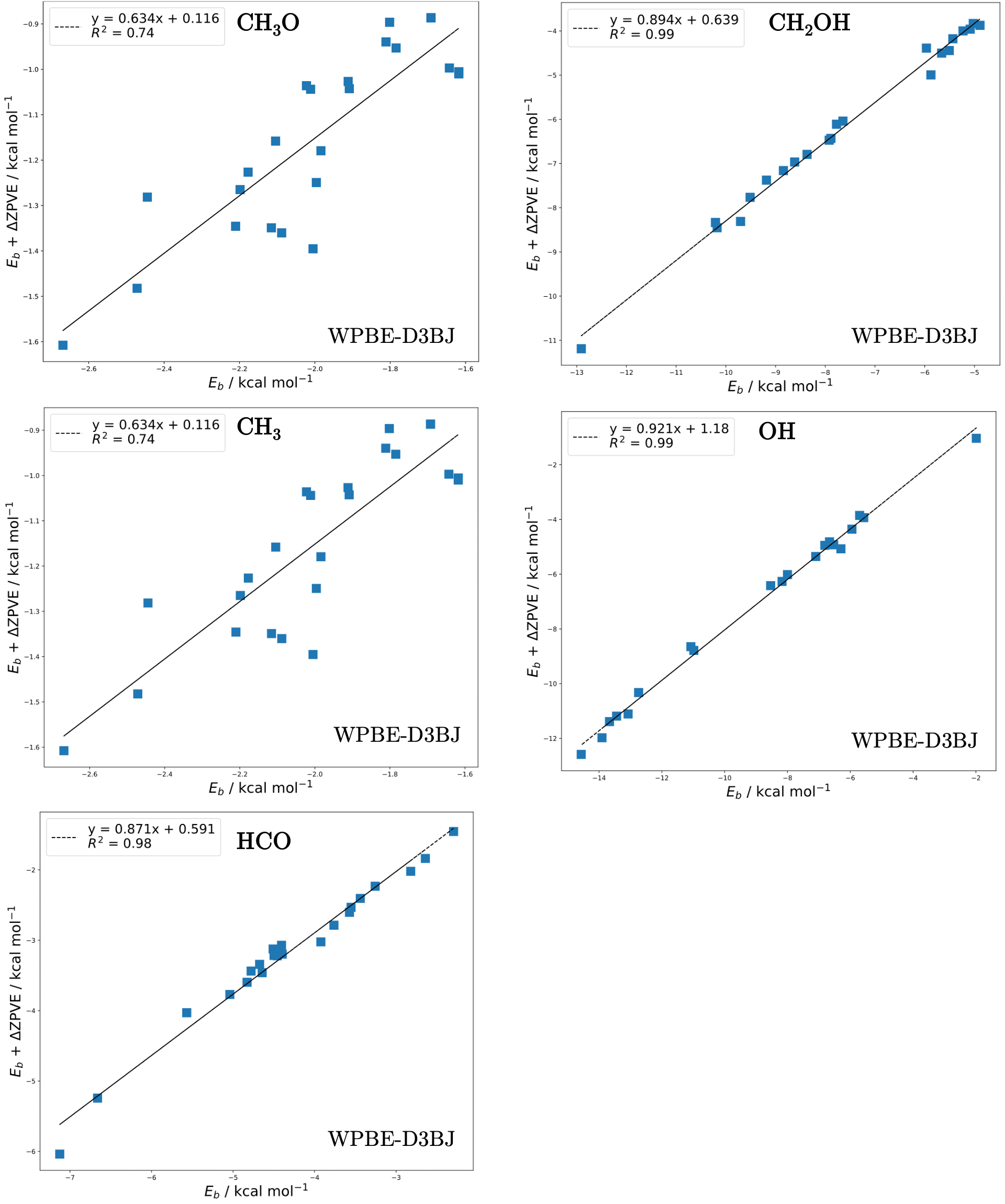}
    \caption{Complete ZPVE models for radicals at the WPBE-D3BJ level of theory.}
    \label{fig:ZPVE_rads}
\end{figure*}

\begin{figure*}
    \centering
    \includegraphics[width=0.75\textwidth]{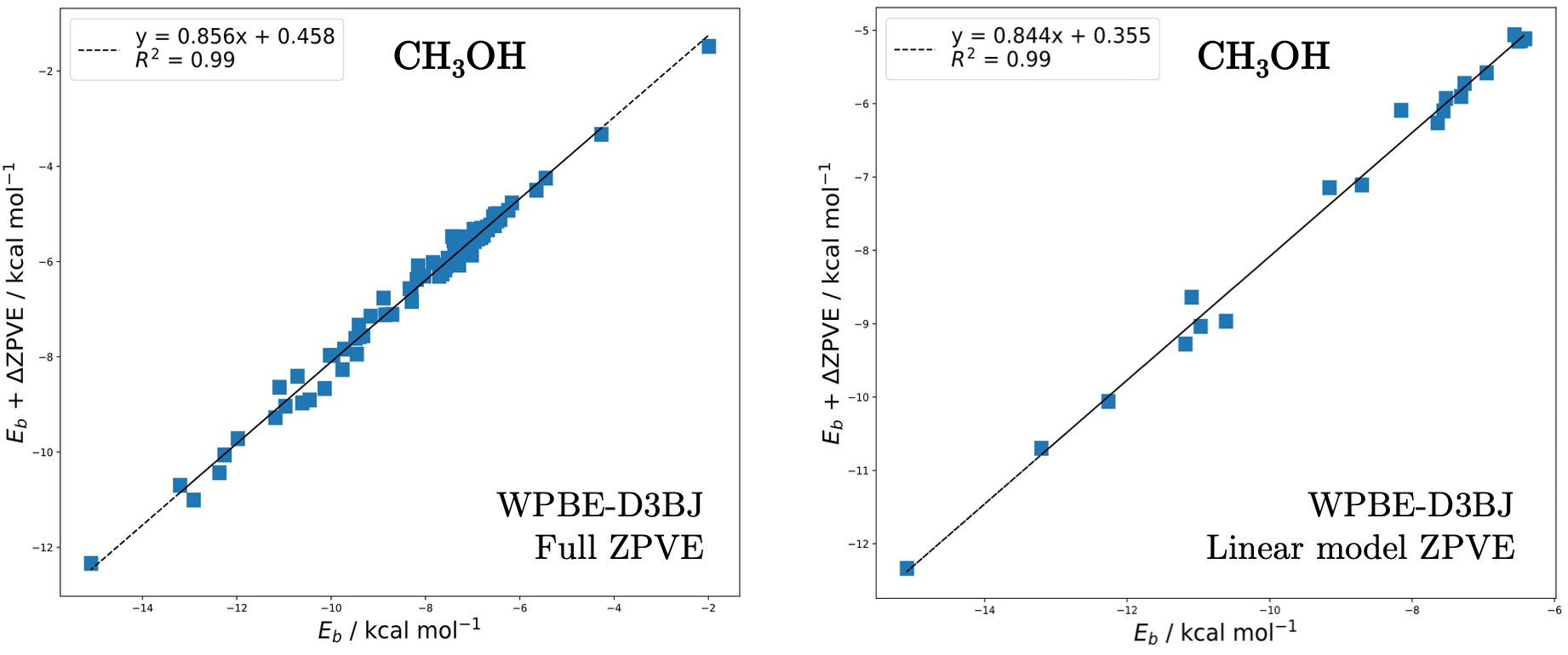}
    \caption{ZPVE models for \ce{CH3OH} . {\em Left:} linear model ZPVE corrected BEs vs BEs  including all hessian matrices computed on all binding sites. {\em Right:} linear model for the ZPVE correction using binding sites on a single cluster.}
    \label{fig:ZPVE_CH3OH}
\end{figure*}

In Figs.~\ref{fig:ZPVE_coms}, \ref{fig:ZPVE_rads}, and \ref{fig:ZPVE_CH3OH}, we show the zero-point vibrational energy (ZPVE) for the closed shell molecules and radicals studied 
in this work. The ZPVE linear model was only applied to the 
radical species for which the Hessian computation on all binding 
sites was prohibitive. Of those, \ce{CH3}, and \ce{CH3O} showed the 
poorest fit; however the uncertainty from the fit of the ZPVE 
is between 0.15 and 0.2 kcal/mol so within the error of the DFT methodology. 
Finally, we evaluated how much the ZPVE linear model changes using all binding sites
versus using the binding sites that are contained only on one ASW
cluster. The linear fit is very similar in both cases, thus 
confirming that building a linear fit with the binding 
sites on one cluster for the radical species is a good 
approximation.

\clearpage

\section{DFT Binding energy benchmark}
\label{sec:bench}

To determine a high quality model chemistry for binding energies, we benchmarked 128 exchange--correlation functionals 
against reference CCSD(T)/CBS energies.  The benchmark set comprises 17
cluster configurations across six molecules: \ce{CH3OH}, \ce{CH4},
\ce{CO}, \ce{H2CO}, \ce{HCO}, and \ce{OH}, with one to three water
molecules in each cluster. 
The reference energies were obtained at the CCSD(T)/CBS level using the
focal-point approach with Dunning's augmented correlation-consistent
basis sets (aug-cc-pVXZ).  The Hartree--Fock energy and MP2 correlation
energy were extrapolated to the complete basis set limit using cardinal
numbers X~=~D, T, Q, while the CCSD(T) correlation contributions
were extrapolated using X~=~D, T.  All DFT single-point energies were
computed on MPWB1K-D3BJ/def2-TZVP optimized geometries using the
def2-TZVP basis set. Open-shell species (\ce{HCO} and \ce{OH}) were
treated with unrestricted Kohn--Sham (UKS).

Table~\ref{tab:BE_closed} presents the top 15 functionals ranked by
mean absolute error (MAE) for closed-shell species (excluding
\ce{CH4}).  Tables~\ref{tab:BE_ch4} and~\ref{tab:BE_uks} report the
rankings for \ce{CH4}--water clusters and open-shell species,
respectively.  

In small cluster systems there are minimal cooperative H-bond effects, however 
on the target ASW$_{12}$ set-of-clusters, cooperative many-body polarization and long-range
electrostatic effects play a significant role. Therefore we selected 
three functionals from the closed-shell ranking
(Table~\ref{tab:BE_closed}) that represent complementary physical
approximations. $\omega$PBE-D3BJ is a range-separated hybrid that 
partitions exchange into short-range semi-local and long-range 
Hartree--Fock components.  The D3BJ empirical dispersion correction provides 
an accurate and computationally inexpensive treatment of dispersion 
interactions. This functional offers a balanced description of both 
hydrogen bonding and dispersion at moderate cost. $\omega$B97X-V,
a range-separated hybrid with the VV10 non-local correlation
functional, was fitted self-consistently into the parameterization.
Unlike the empirical D3 correction, the VV10 kernel captures
dispersion through the electron density itself, providing a more
physically motivated treatment of non-local correlation effects
that become increasingly important in extended hydrogen-bond
networks. Finally, M06-HF is a meta-GGA hybrid with 100\% Hartree--Fock exchange.  
While M06-2X performs slightly better on the small benchmark clusters, 
M06-HF was selected for the larger set-of-clusters because full exact
exchange eliminates the delocalization error inherent in semi-local 
exchange, which accumulates across the cooperative hydrogen-bond network 
of larger water clusters.
Together, these three functionals span the dominant sources of error in
DFT descriptions of non-covalent interactions: short- versus long-range
exchange partitioning ($\omega$PBE-D3BJ), density-based versus empirical
dispersion ($\omega$B97X-V versus \ $\omega$PBE-D3BJ), and delocalization
error (M06-HF).  Any systematic deviation shared by all three is
unlikely to arise from a single methodological artifact, strengthening
confidence in the resulting binding energies for the ASW$_{12}$ water clusters.

\begin{deluxetable*}{rlrrrc}
\tablecaption{Top 15 DFT functionals for the BEs of closed-shell
molecule--water clusters excluding \ce{CH4} (i.e., including \ce{CH3OH}, \ce{CO},
\ce{H2CO}; 11 configurations).  The MAE (mean asbolute error), RMSE (root-mean-square error), and Max AE (absolute error) are listed in kcal\,mol$^{-1}$.  $N$ is the number of systems for which the
functional was evaluated. \label{tab:BE_closed}}
\tablehead{
  \colhead{Rank} & \colhead{Functional} & \colhead{MAE} & \colhead{RMSE} & \colhead{Max AE} & \colhead{$N$}
}
\startdata
    1 & $\omega$B97M-V            & 0.185 & 0.283 & 0.631 & 11 \\
    2 & $\omega$PBE-D3(MBJ)       & 0.202 & 0.282 & 0.613 & 11 \\
    3 & $\omega$B97X-V            & 0.212 & 0.279 & 0.694 & 11 \\
    4 & $\omega$B97M-D3BJ         & 0.231 & 0.297 & 0.677 & 11 \\
    5 & $\omega$PBE-D3BJ          & 0.260 & 0.323 & 0.654 & 11 \\
    6 & SCAN0                     & 0.275 & 0.401 & 1.070 & 11 \\
    7 & M05-2X                    & 0.275 & 0.336 & 0.630 & 11 \\
    8 & M06-2X                    & 0.280 & 0.394 & 0.871 & 11 \\
    9 & B1LYP-D3BJ                & 0.282 & 0.368 & 0.665 & 11 \\
   10 & M06-HF                    & 0.295 & 0.382 & 0.904 & 11 \\
   11 & $\omega$B97X-D            & 0.300 & 0.352 & 0.689 & 11 \\
   12 & SOGGA11-X-D3BJ            & 0.312 & 0.364 & 0.748 & 11 \\
   13 & M11-D3BJ                  & 0.326 & 0.424 & 0.843 & 11 \\
   14 & PWB6K-D3BJ                & 0.326 & 0.380 & 0.792 & 11 \\
   15 & revPBE0-D3BJ              & 0.336 & 0.424 & 0.876 & 11 \\
\enddata
\end{deluxetable*}

\begin{deluxetable*}{rlrrrc}
\tablecaption{Top 15 DFT functionals for the BEs of
\ce{CH4}--water clusters (2 configurations).  All values are listed in
kcal\,mol$^{-1}$. \label{tab:BE_ch4}}
\tablehead{
  \colhead{Rank} & \colhead{Functional} & \colhead{MAE} & \colhead{RMSE} & \colhead{Max AE} & \colhead{$N$}
}
\startdata
    1 & $\omega$B97M-V            & 0.016 & 0.022 & 0.031 & 2 \\
    2 & B97-2-D3BJ                & 0.052 & 0.064 & 0.090 & 2 \\
    3 & TPSSh-D3BJ                & 0.059 & 0.084 & 0.118 & 2 \\
    4 & revPBE0-D3BJ              & 0.071 & 0.099 & 0.140 & 2 \\
    5 & B3LYP-D3BJ                & 0.073 & 0.073 & 0.074 & 2 \\
    6 & B1LYP-D3BJ                & 0.079 & 0.079 & 0.082 & 2 \\
    7 & HF+D                      & 0.080 & 0.103 & 0.145 & 2 \\
    8 & $\omega$PBE-D3BJ          & 0.082 & 0.111 & 0.157 & 2 \\
    9 & M08-HX                    & 0.083 & 0.086 & 0.105 & 2 \\
   10 & PWB6K                     & 0.083 & 0.103 & 0.144 & 2 \\
   11 & N12-SX-D3BJ               & 0.085 & 0.088 & 0.109 & 2 \\
   12 & CAM-B3LYP-D3BJ            & 0.088 & 0.099 & 0.134 & 2 \\
   13 & $\omega$B97X-V            & 0.093 & 0.106 & 0.144 & 2 \\
   14 & revTPSSh-D3BJ             & 0.099 & 0.113 & 0.154 & 2 \\
   15 & HSE06-D3BJ                & 0.105 & 0.111 & 0.143 & 2 \\
\enddata
\end{deluxetable*}

\begin{deluxetable*}{rlrrrc}
\tablecaption{Top 15 DFT functionals for the BEs of open-shell
molecule--water clusters (\ce{HCO}, \ce{OH}; 4 configurations,
unrestricted Kohn--Sham).  All values are listed in
kcal\,mol$^{-1}$. \label{tab:BE_uks}}
\tablehead{
  \colhead{Rank} & \colhead{Functional} & \colhead{MAE} & \colhead{RMSE} & \colhead{Max AE} & \colhead{$N$}
}
\startdata
    1 & MPWB1K-D3BJ               & 0.188 & 0.231 & 0.371 & 4 \\
    2 & $\omega$PBE-D3BJ          & 0.193 & 0.225 & 0.346 & 4 \\
    3 & M05                       & 0.225 & 0.322 & 0.615 & 4 \\
    4 & PWB6K                     & 0.227 & 0.328 & 0.637 & 4 \\
    5 & revPBE0-D3BJ              & 0.243 & 0.342 & 0.568 & 4 \\
    6 & B97-2-D3BJ                & 0.251 & 0.287 & 0.467 & 4 \\
    7 & MGGA\_MS2h                & 0.253 & 0.291 & 0.392 & 4 \\
    8 & PW6B95-D3BJ               & 0.256 & 0.266 & 0.347 & 4 \\
    9 & $\omega$PBE-D3(MBJ)       & 0.260 & 0.310 & 0.500 & 4 \\
   10 & PTPSS-D3BJ                & 0.271 & 0.303 & 0.401 & 4 \\
   11 & $\omega$B97X-D            & 0.272 & 0.315 & 0.407 & 4 \\
   12 & $\omega$B97M-V            & 0.273 & 0.344 & 0.504 & 4 \\
   13 & MPW1B95-D3BJ              & 0.274 & 0.317 & 0.481 & 4 \\
   14 & MN12-SX-D3BJ              & 0.274 & 0.344 & 0.617 & 4 \\
   15 & BMK-D3BJ                  & 0.286 & 0.300 & 0.414 & 4 \\
\enddata
\end{deluxetable*}

\end{document}